\RequirePackage{fix-cm}
\documentclass[smallextended]{svjour3}       % onecolumn (second format)
\smartqed  % flush right qed marks, e.g. at end of proof
\usepackage{gensymb}
\usepackage{cite}
\usepackage{lineno}
\usepackage{graphicx}
\usepackage{amsmath}
\usepackage{epsfig}
\usepackage{color}
\usepackage[left=1in, right=1in, top=0.75in, bottom=0.75in, includefoot]{geometry}
\newcommand{\ve}{\varepsilon}
\renewcommand{\Re}{\mbox{Re}}
\newcommand{\We}{\mbox{We}}
\newcommand{\Oh}{\mbox{Oh}}

  \def\Xint#1{\mathchoice
   {\XXint\displaystyle\textstyle{#1}}%
   {\XXint\textstyle\scriptstyle{#1}}%
   {\XXint\scriptstyle\scriptscriptstyle{#1}}%
   {\XXint\scriptscriptstyle\scriptscriptstyle{#1}}%
   \!\int}
\def\XXint#1#2#3{{\setbox0=\hbox{$#1{#2#3}{\int}$}
     \vcenter{\hbox{$#2#3$}}\kern-.5\wd0}}

\def\dashint{\Xint-}

\begin{document}

\title{Introducing pre-impact air-cushioning effects into the Wagner model of impact theory%\thanks{Grants or other notes
%about the article that should go on the front page should be
%placed here. General acknowledgments should be placed at the end of the article.}
}
\subtitle{}

%\titlerunning{Short form of title}        % if too long for running head

\author{Matthew R. Moore}

%\authorrunning{Short form of author list} % if too long for running head

\institute{M. R. Moore\at
              Mathematical Institute, Radcliffe Observatory Quarter, St. Giles, Oxford, OX2 6GG, UK. \\
              \email{moorem@maths.ox.ac.uk}           %  \\
%             \emph{Present address:} of F. Author  %  if needed
%            \and
%            S. Author \at
%               second address
}

\date{Received: date / Accepted: date}
% The correct dates will be entered by the editor

\maketitle

\begin{abstract}
In this analysis, we consider the effects of non-quiescent initial conditions driven by pre-impact air-water interactions on the classical Wagner model of impact theory. We consider the problem of a rigid, solid impactor moving vertically towards a liquid pool. Prior to impact, viscous forces in the air act to deform the liquid free surface, inducing a flow in the pool. These interactions are then incorporated as initial conditions in the post-impact analysis. We derive expressions for the size of the effective contact set, the leading-order pressure and force on the impactor, and the speed and thickness of the jet at its base. In all cases, we show that the effect of the pre-impact behaviour is to cushion the impactor, reducing the size of the effective contact set and, hence, the force on the impactor. Small- and large-time asymptotic solutions are derived for general power-law impactors, and we show that the effects of the air die away as the impact progresses, so that we approach the classical Wagner solution.

\keywords{Water entry \and Air cushioning \and Ship slamming}
% \PACS{PACS code1 \and PACS code2 \and more}
% \subclass{MSC code1 \and MSC code2 \and more}
\end{abstract}

% \linenumbers

\section{Introduction} \label{sec:Introduction}

Wagner theory is one of the success stories for practical asymptotics. Named after the author of the seminal paper in the field, Wagner theory is an inviscid fluid model for the water-entry of a rigid, solid body into a quiescent pool of liquid, mimicking the landing of a sea-plane or the slamming of a ship over a wave in an ocean \cite{Wagner1932}. Central to both Wagner's model and earlier work by von K\'{a}rm\'{a}n \cite{VonKarman1929} is the idea that, since most impactor profiles relevant to naval architecture are relatively flat, the penetration depth shortly after entry is much smaller than the horizontal wetted length of the body. In technical terms, the impactor is assumed to have small \emph{deadrise angle}. This allows the modeller to approximate a two-dimensional cross section of the entering body as a flat plate. This, coupled with the typically large Reynolds, Weber and Froude numbers involved in water entry, enables the use of powerful complex analytical techniques to solve the resulting ideal fluid model. 

What set Wagner's model ahead of earlier work was his method for determining the size of the approximate flat plate, which is not known at the outset --- indeed, the linearized problem is an example of codimension-two free boundary problem, \cite{Howison1997}. By conservation of mass, Wagner argued that the free surface of the liquid must rise to meet the body profile at its edges, providing a condition that enables the size of the plate to be determined. This requirement --- called the \emph{Wagner condition} --- can be rigorously derived as a matching condition between the flat-plate model and the local Helmholtz flow close to the plate tips, where the free surface turns over into a splash jet, \cite{Howison1991}. This has led to the term \emph{turnover points} to describe the ends of the effective flat plate in Wagner theory. Significant efforts have been made to formalize Wagner theory using an asymptotic treatment in terms of the small deadrise angle (or equivalently, the small impact timescale), see for example, \cite{Pukhnachov1981, Korobkin1985, Cointe1987, Howison1991}. 

The asymptotic predictions have been shown to be favourable when compared to experiments and simulations. In \cite{Howison1991}, comparisons between the leading-order composite pressure profile at different points on the impactor and experimental data from drop tests of a parabolic profile in \cite{Nethercote1986} were shown to be in good qualitative agreement. The comparison can be drastically improved by going to second-order in the theory, \cite{Oliver2007}. For a wedge-shaped body, the error between the Wagner model and numerical data in \cite{Zhao1993} for the size of the effective wetted length of the wedge (i.e. the flat plate length) was less than 10\% for bodies of deadrise angle as large as 30\degree. For parabolic impactors, \cite{Oliver2007} shows that the second-order theory increases the time over which Wagner theory is within 10\% of experimental data in \cite{Campbell1980} by two orders of magnitude. 

Recently, models inspired by Wagner theory have also been used in problems of droplet impact, where, although the scales involved are much smaller, the dimensionless numbers typically remain large, and the mathematics is very similar. Droplet impacts have important applications in, for example, inkjet printing \cite{Derby2010}, soil erosion \cite{Smith1990}, and droplet-size distribution in aerosols \cite{Moreira2010}, the latter of which can be a crucial factor in understanding the transmission of viral diseases \cite{Morawska2005, Yang2007}, or even the dispersal of marine oil slicks due to breaking waves \cite{Murphy2015}. Inviscid Wagner models have been shown to provide excellent predictions of general properties of the impact such as the effective wetted region and the speed of the splash jet (or ejecta), \cite{Philippi2016, Cimpeanu2018}, although they do not fare so well in capturing the finer details of droplet impact, in particular regarding the evolution and eventual break-up of the ejecta, where further physical effects become relevant \cite{Thoraval2012, Gordillo2015, Moore2018}.

These successes have inspired a number of extensions to the classical theory, including the effect of an oblique component of impact velocity \cite{Korobkin1988, Judge2004, Moore2012, Moore2013b}, droplet impact onto soft substrates \cite{Howland2016} and porous media \cite{Hicks2017}, or impacts involving elastic bodies or plates \cite{Khabakhpasheva2003, Korobkin2016, Pegg2018}. 

However, of particular interest to the present study is the role of the surrounding air in an impact. While the aforementioned comparisons between Wagner theory and experimental data in \cite{Howison1991} were qualitatively very good, there were certain discrepancies between the results, particularly close to the keel of the impactor. In the experimental data of \cite{Nethercote1986}, oscillations were observed in the pressure on the impactor close to the keel. This is attributed to the presence of an entrapped air bubble, which had been observed previously in experiments of the impact of flat-bottomed bodies in \cite{Fujita1954, Chuang1966}. The presence of the air layer can significantly reduce the loads on the impacting body, \cite{Chuang1966, Hagiwara1974, Huera2011}, which has lead to the term \emph{air cushioning} to describe this effect. 

In brief, the air bubble is entrapped as follows. Prior to impact, as the solid body approaches the liquid pool (or vice versa), the air layer between the impactor and the pool is forced out of the way, which deforms the free surface of the liquid and induces a liquid flow as a result. The free surface rises to meet the impactor, and the first touchdown of the solid typically occurs away from the midline of the impactor, entrapping a volume of air. This air then contracts into a bubble \cite{Thoroddsen2005, Thoroddsen2010, Marston2011}, which can be a major cause for concern in, for example, the printing of polymetric circuits \cite{Sirringhaus2000}.

While the earliest attempts to model this effect considered an inviscid model for the air flow, \cite{Verhagen1967, Wilson1991}, it is clear that the air layer becomes sufficiently thin for viscous forces to dominate. A number of recent studies have demonstrated that a viscous air-inviscid fluid coupled model produces this air-cushioning effect \cite{Smith2003, Purvis2004, Mandre2009, Mani2010, Hicks2010, Hicks2013}, and recent comparisons between the theoretical predictions and experimental results of the size of the entrapped air bubble have been shown to be excellent \cite{Hicks2012, Bouwhuis2015}.

However, attempts to incorporate air-cushioning back into Wagner theory have been few and far between, with the majority concentrating on the effect of air cushioning post-impact \cite{Purvis2004, Moore2013, Moore2014}, rather than its much larger effect before touchdown. An excellent recent study \cite{Ross2019} considers the differences between a classical Wagner model in the absence of air cushioning to the viscous air-inviscid liquid pre-impact model. At touchdown in the air cushioning model, there was broad qualitative agreement between the Wagner predictions for the free surface profile, the velocity potential on the free surface and the turnover point location (equated to the point of minimum air layer thickness in the pre-impact regime), particularly when the effects of surface tension are negligible. While this is an encouraging sign for the classical Wagner model, there were nonetheless differences between the two models, and it is clear that the pre-impact air effects not only play a role on the force felt by the impactor --- as evidenced also by the aforementioned experiments --- but also on the general post-impact dynamics. There is an induced flow in the liquid, so the assumption of an initially quiescent liquid pool in the Wagner model is unrealistic, and it is not clear how incorporating this effect alters the structure and the predictions of Wagner theory.

It is the goal of this analysis to investigate these effects in detail. We will build in the pre-impact model of \cite{Ross2019} as initial conditions to a Wagner model in which touchdown occurs away from the centreline of the impacting body. Exploiting results of the air flow at touchdown and the relatively long timescale of bubble collapse, we will neglect the collapse of the air pocket into numerous bubbles in modelling the post-impact dynamics. We will solve the resulting model and extract crucial predictions of the post-impact dynamics, such as the location of the turnover points and the slamming force on the impactor. Since it includes the pre-impact air behaviour, this model is a significant step forward from classical Wagner theory and it will become clear that the air plays an important role in the post-impact solution at early and intermediate times, although the effect dies away as we move further from touchdown.

The structure of the analysis is as follows. In \textsection \ref{sec:PreImpact}, we will formulate the problem, and briefly outline the pre-impact dynamics derived in \cite{Ross2019}, as they are central to the post-impact analysis we move onto in \textsection \ref{sec:PostImpact}. We will illustrate the results of the post-impact analysis in \textsection \ref{sec:PostImpactExample}, in particular looking at the impact of a parabolic body profile. Finally, we shall conclude by summarizing our results, discuss applications and suggest future avenues of investigation in \textsection \ref{sec:Summary}.

%%%%%%%%%%%%%%%%%%%%%%%%%%%%%%%%%%%%%%%%%%%%%%%%%%%%%%%%%%%%%%%%%%%%%%%%%%%%
%%%%%%%%%%%%%%%%%%%%%%%%%%%%%%%%%%%%%%%%%%%%%%%%%%%%%%%%%%%%%%%%%%%%%%%%%%%%
%%%%%%%%%%%%%%%%%%%%%%%%%%%%%%%% PRE-IMPACT %%%%%%%%%%%%%%%%%%%%%%%%%%%%%%%%
%%%%%%%%%%%%%%%%%%%%%%%%%%%%%%%%%%%%%%%%%%%%%%%%%%%%%%%%%%%%%%%%%%%%%%%%%%%%
%%%%%%%%%%%%%%%%%%%%%%%%%%%%%%%%%%%%%%%%%%%%%%%%%%%%%%%%%%%%%%%%%%%%%%%%%%%%

\section{Pre-impact model} \label{sec:PreImpact}

We shall first consider the dynamics before impact in which a large, two-dimensional solid body hurtles towards an initially-quiescent liquid pool whose depth is sufficiently large that we can take to be infinite. The region that is not occupied by the impactor or the liquid pool is filled with air.  In what follows, we are following the analysis of \cite{Ross2019} and we shall only outline the key results here, with the reader directed to that paper for a more detailed discussion of the coupling.

The body is assumed to be symmetric and, when it is sufficiently far from the liquid pool, the liquid free surface is flat. We shall take Cartesian axes $(x^{*},y^{*})$ centred on the point where line of symmetry meets the initially flat liquid surface. Here and hereafter an asterisk denotes a dimensional variable. The body moves vertically downwards with a constant velocity $-V^{*}\textbf{j}$. We shall assume that the impactor has a small \emph{deadrise angle}, which means that it is essentially flat close to the region of impact. The body profile is taken to be
\begin{linenomath}
\begin{equation}
 y^{*} = \ve^{2}R^{*}f\left(\frac{x^{*}}{\ve R^{*}}\right) - V^{*}t^{*},
 \label{eqn:BodyProfile}
\end{equation}
\end{linenomath}
where $0<\ve\ll1$ and $R^{*}$ is a typical radius of curvature of the impactor.

Both the air and the liquid are taken to be Newtonian, incompressible fluids, with densities $\rho_{i}^{*}$ and viscosities $\mu_{i}^{*}$, where $i = a$ in the air and $i = l$ in the liquid. The air-water surface tension coefficient is $\sigma^{*}$ and acceleration due to gravity is $g^{*}$. The pertinent physical parameters measuring the importance of surface tension and gravity are the Weber and Froude numbers
\begin{linenomath}
\begin{equation}
 \mbox{We} = \frac{\rho_{l}^{*}R^{*}V^{*2}}{\sigma^{*}}, \quad \mbox{Fr}^{2} = \frac{V^{*2}}{g^{*}R^{*}}.
\end{equation}
\end{linenomath}
For typical solid-liquid impacts of interest, these numbers are very large, so for the rest of this analysis we shall neglect surface tension and gravity, \cite{Howison1991, Ross2019}. We shall denote the fluid velocities and pressures by $\textbf{u}_{i}^{*} = (u_{i}^{*},v_{i}^{*})$ and $p_{i}^{*}$ respectively.

As shown by \cite{Ross2019} and seen from (\ref{eqn:BodyProfile}), the body starts to interact with the pool when it is an $O(\ve^{2}R^{*})$ distance from the free surface. Since the body has small deadrise angle, these air-liquid interactions take place over a horizontal lengthscale of $O(\ve R^{*})$ and a timescale of $O(\ve^{2}R^{*}/V^{*})$. The impactor displaces the air in the region between itself and the liquid, inducing a vertical velocity of $O(V^{*})$ and, by conservation of mass, a horizontal velocity of $O(V^{*}/\ve)$. The air motion displaces the liquid free surface by a distance of $O(\ve^{2}R^{*})$, which in turn induces a velocity within the liquid pool of $O(V^{*})$.

Since we shall assume that the Reynolds number
\begin{linenomath}
\begin{equation}
 \mbox{Re} = \frac{\rho_{l}^{*}R^{*}V^{*}}{\mu_{l}^{*}}
\end{equation}
\end{linenomath}
in the liquid pool is large, to maintain a balance in the momentum equations, the liquid pressure is perturbed by $O(\rho_{l}^{*} V^{*2}/\ve)$. Hence, to retain a balance in the normal stress condition on the free surface, the air pressure is also perturbed by $O(\rho_{l}^{*}V^{*2}/\ve)$. Thus, finally, to retain a balance in the horizontal momentum equation in the air layer, we therefore find that interactions occur over the lengthscale
\begin{linenomath}
\begin{equation}
 \ve^{2}R^{*} = \left(\frac{\mu_{a}^{*}}{\rho_{l}^{*}R^{*}V^{*}}\right)^{2/3}R^{*},
 \label{eqn:epsilon}
\end{equation} 
\end{linenomath}
as in \cite{Ross2019}.

Hence, we may nondimensionalize the problem by setting
\begin{linenomath}
\begin{equation}
 \left(x^{*},y^{*},h^{*},t^{*},u_{a}^{*},v_{a}^{*},p_{a}^{*}\right) = \left(\ve R^{*} x, \ve^{2}R^{*}\bar{y},\ve^{2}R^{*} h,\ve^{2}R^{*} t/V^{*}, V^{*}u_{a}/\ve, V^{*}v_{a}, \rho_{l}^{*}V^{*2}p_{a}/\ve \right)
\end{equation}
\end{linenomath}
in the air, and
\begin{linenomath}
\begin{equation}
\left(x^{*},y^{*},h^{*},t^{*},u_{l}^{*},v_{l}^{*},p_{l}^{*}\right) = \left(\ve R^{*} x, \ve R^{*}y,\ve^{2}R^{*} h,\ve^{2}R^{*} t/V^{*}, V^{*}u_{l}, V^{*}v_{l}, \rho_{l}^{*}V^{*2}p_{l}/\ve \right)
\label{eqn:NonDimLiquid}
\end{equation}
\end{linenomath}
in the liquid, where $h^{*}$ denotes the air-liquid interface. Then, under the assumptions that
\begin{linenomath}
\begin{equation}
  \frac{\rho_{a}^{*}}{\rho_{l}^{*}} \ll \ve, \quad \frac{1}{\Re}\ll1, \quad \frac{\mu_{a}^{*}}{\mu_{l}^{*}} \gg \ve^{2},
\end{equation}
\end{linenomath}
we are able to neglect, respectively, inertia in the air, viscosity in the liquid and air shear on the free surface. These are all reasonable assumptions in typical impact problems, \cite{Smith2003, Ross2019}, and we shall neglect these forthwith.

The leading-order problem in the air is thus given by
\refstepcounter{equation}
\begin{linenomath}
$$
 \frac{\partial u_{a}}{\partial x} + \frac{\partial v_{a}}{\partial \bar{y}} = 0, \quad 0 = -\frac{\partial p_{a}}{\partial x} + \frac{\partial^{2}u_{a}}{\partial \bar{y}^{2}}, \quad 0 = -\frac{\partial p_{a}}{\partial \bar{y}},
  \eqno{(\theequation{\mathit{a},\mathit{b}},\mathit{c})}
  \label{eqn:LeadingOrderAir}
$$
\end{linenomath}
while in the liquid we must have
\refstepcounter{equation}
\begin{linenomath}
$$
 \frac{\partial u_{l}}{\partial x} + \frac{\partial v_{l}}{\partial y} = 0, \quad \frac{\partial u_{l}}{\partial t} = -\frac{\partial p_{l}}{\partial x}, \quad \frac{\partial v_{l}}{\partial t} = -\frac{\partial p_{l}}{\partial y}.
   \eqno{(\theequation{\mathit{a},\mathit{b}},\mathit{c})}
 \label{eqn:LeadingOrderLiquid}
$$
\end{linenomath}
On the solid body, the no-slip, no-flux conditions reduce to
\refstepcounter{equation}
\begin{linenomath}
$$
 u_{a} = 0, \quad v_{a} = -1 \quad \mbox{on} \quad \bar{y} = f(x) - t,
 \eqno{(\theequation{\mathit{a},\mathit{b}})}
 \label{eqn:NSNF}
$$
\end{linenomath}
while on the air-liquid interface, we have continuity of velocity, the kinematic condition and continuity of normal stress, which respectively reduce to
\refstepcounter{equation}
\begin{linenomath}
$$
 u_{a} = 0, \quad v_{a} = v_{l}(x,\ve h, t), \quad v_{a} = \frac{\partial h}{\partial t}, \quad p_{a} = p_{l}(x,\ve h, t) \quad \mbox{on} \quad \bar{y} = h.
   \eqno{(\theequation{\mathit{a},\mathit{b}},\mathit{c}, \mathit{d})}
 \label{eqn:DBC}
$$
\end{linenomath}
Far from the impact, the flow in both fluids dies away, so we must have
\refstepcounter{equation}
\begin{linenomath}
$$
 u_{a} \rightarrow 0 \quad \mbox{as} \quad |x|\rightarrow\infty, \quad \textbf{u}_{l} \rightarrow 0 \quad \mbox{as} \quad x^{2}+y^{2}\rightarrow\infty, \quad h \rightarrow 0 \quad \mbox{as} \quad |x|\rightarrow\infty.
   \eqno{(\theequation{\mathit{a},\mathit{b}},\mathit{c})}
$$
\end{linenomath}
% \begin{linenomath}
% \begin{alignat}{4}
%  u_{a} & \, \rightarrow && \; 0 && \quad \mbox{as} && \quad |x|\rightarrow\infty, \\
%  \textbf{u}_{l} & \, \rightarrow && \; 0 && \quad \mbox{as} && \quad x^{2}+y^{2}\rightarrow\infty, \\
%  h & \, \rightarrow && \; 0 && \quad \mbox{as} && \quad |x|\rightarrow\infty.
% \end{alignat}
% \end{linenomath}
Finally, we shall assume that initially the free surface is flat and the fluids are undisturbed, so that
\refstepcounter{equation}
\begin{linenomath}
$$
 \textbf{u}_{a} \rightarrow 0 \quad \mbox{as} \quad t\rightarrow-\infty \quad \mbox{for} \quad \bar{y} = O(1), \quad \textbf{u}_{l} \rightarrow  0 \quad \mbox{as} \quad t\rightarrow-\infty, \quad h \rightarrow 0 \quad \mbox{as} \quad t\rightarrow-\infty.
   \eqno{(\theequation{\mathit{a},\mathit{b}},\mathit{c})}
$$
\end{linenomath}
% \begin{linenomath}
% \begin{alignat}{4}
% \textbf{u}_{a} & \, \rightarrow && \; 0 && \quad \mbox{as} && \quad t\rightarrow-\infty \quad \mbox{for} \quad \bar{y} = O(1), \\
%  \textbf{u}_{l} &\, \rightarrow && \; 0 && \quad \mbox{as} && \quad t\rightarrow-\infty, \\
%  h & \, \rightarrow && \; 0 && \quad \mbox{as} && \quad t\rightarrow-\infty.
% \end{alignat}
% \end{linenomath}

From (\ref{eqn:LeadingOrderAir}c) and (\ref{eqn:DBC}d) it is clear that to leading order $p_{a} = p_{l}(x,0,t) := P(x,t)$ throughout the air layer. Thus, from (\ref{eqn:LeadingOrderAir}b), (\ref{eqn:NSNF}a) and (\ref{eqn:DBC}a), the leading-order air velocity has the Poiseuille profile
\begin{linenomath}
\begin{equation}
 u_{a} = \frac{1}{2}\frac{\partial P}{\partial x}\left[(\bar{y}-h)^{2} - (f-t-h)(\bar{y}-h)\right],
\end{equation}
\end{linenomath}
so that integrating the continuity equation (\ref{eqn:LeadingOrderAir}a) across the layer and applying the no-flux condition (\ref{eqn:NSNF}b) and kinematic condition (\ref{eqn:DBC}c) yields the thin-film equation
\begin{linenomath}
\begin{equation}
 0 = \frac{\partial}{\partial t}\left(f-t-h\right) - \frac{1}{12}\frac{\partial}{\partial x} \left(Q(f-t-h)^{3}\right), \label{eqn:PreImpact1}
\end{equation}
\end{linenomath}
as in \cite{Ross2019}. We note that here we have introduced the notation $Q(x,t) = P_{x}(x,t)$ for convenience, and we shall retain this henceforth.

In the liquid, we shall assume that the flow is initially irrotational; as it is quiescent as $t\rightarrow-\infty$, this is eminently sensible. Thus, by Kelvin's theorem, we have $\textbf{u}_{l} = \nabla\phi$, such that $\phi$ satisfies $\nabla^{2}\phi = 0$ in $y < 0$. Moreover, the velocity potential and pressure are related by the linearized Bernoulli equation $p_{l} = -\partial\phi_{l}/\partial t$. Thence, after linearizing the boundary conditions (\ref{eqn:DBC}) onto the undisturbed waterline, the pressure satisfies the half-plane Laplace problem:
\refstepcounter{equation}
\begin{linenomath}
$$
 \nabla^{2}p_{l} = 0 \quad \mbox{in} \quad y < 0, \quad p_{l} = P(x,t) \quad \mbox{on} \quad y = 0, \quad \frac{\partial p_{l}}{\partial y} = -\frac{\partial^{2}h}{\partial t^{2}} \quad \mbox{on} \quad y = 0. 
   \eqno{(\theequation{\mathit{a},\mathit{b}},\mathit{c})}
$$
\end{linenomath}
% \begin{linenomath}
% \begin{alignat}{4}
%  \nabla^{2}p_{l} & \, = && \, 0 && \, \mbox{in} && \, y < 0, \\
%  p_{l} & \, = && \, P(x,t) && \, \mbox{on} && \, y = 0, \\
%  \frac{\partial p_{l}}{\partial y} & \, = && \, \tcb{-\frac{\partial^{2}h}{\partial t^{2}}} && \, \mbox{on} && \, y = 0. 
% \end{alignat}
% \end{linenomath}
The appropriate Dirichlet-to-Neumann map relating the free surface profile, $h(x,t)$, and the liquid pressure gradient on the interface, $Q(x,t)$, is then found to be
\begin{linenomath}
\begin{equation}
 \frac{\partial^{2}h}{\partial t^{2}} = \frac{1}{\pi}\dashint_{-\infty}^{\infty} \frac{Q(\xi,t)}{\xi-x}\,\mbox{d}\xi, \label{eqn:PreImpact2}
\end{equation}
\end{linenomath}
for $-\infty<x<\infty$.

Hence, the pre-impact dynamics are described by the coupled singular integro-differential system (\ref{eqn:PreImpact1}), (\ref{eqn:PreImpact2}) subject to suitable initial and far-field conditions. The system is readily solved numerically by adapting the methodology of \cite{Hicks2017}, in which $Q$ and $h$ are represented by complex Fourier series, which allows (\ref{eqn:PreImpact2}) to be solved as a system of ordinary differential equations for the time-dependent Fourier coefficients. The lubrication equation (\ref{eqn:PreImpact1}) can then be solved as an ordinary differential equation for $Q$ given $h$. This process is iterated at each time station until a desired convergence is achieved. The code terminates when the minimum thickness between the body and the pool reaches a certain tolerance to indicate touchdown. There is a delicate balance between the convergence rate of the iterative procedure and the closeness to touchdown: for the simulations presented in this paper, since we are concentrating on the post-impact behaviour rather than a detailed analysis of touchdown, we terminated the code once this thickness was less than $10^{-2}$ dimensionless units. 

Since the time domain extends to $-\infty$, the numerical methodology is quite sensitive to the correct early-time behaviour when the impactor is far from the pool, so knowledge of the asymptotic solution for large negative time is necessary to retain good accuracy in the simulations, and we discuss this in \textsection \ref{sec:PreImpactAsymptotics}. 

As shown in detail in \cite{Ross2019}, as time increases and the impactor approaches the free surface, there is a build up of pressure in the air layer, which leads to symmetric deformation of the free surface. When the body is sufficiently smooth (i.e. smoother than a wedge), the free surface depresses about $x = 0$, and rises up at $x = \pm d(t)$, which leads to a further increase in pressure at these rises, where the air layer is thinnest. In the limit in which surface tension is negligible, the solution breaks down at a time $t = t_{c}$, when there is simultaneous touchdown at $x = \pm d_{c} = \pm d(t_{c})$. The free surface, its velocity $v_{c}(x)$ and the velocity potential on $y = 0$ at touchdown are then given by
\begin{linenomath}
\begin{equation}
 h_{c}(x) = h(x,t_{c}), \quad v_{c}(x) = \frac{\partial\phi_{c}}{\partial y}(x,0,t_{c}), \quad \phi_{c}(x,y) = \phi(x,y,t_{c}),
 \label{eqn:InitialConditions}
\end{equation}
\end{linenomath}
where the velocity potential is calculated from Bernoulli's equation subject to the initial condition that $\phi\rightarrow0$ as $t\rightarrow-\infty$. 

\begin{figure}[t!]
\centering \scalebox{0.45}{\epsfig{file=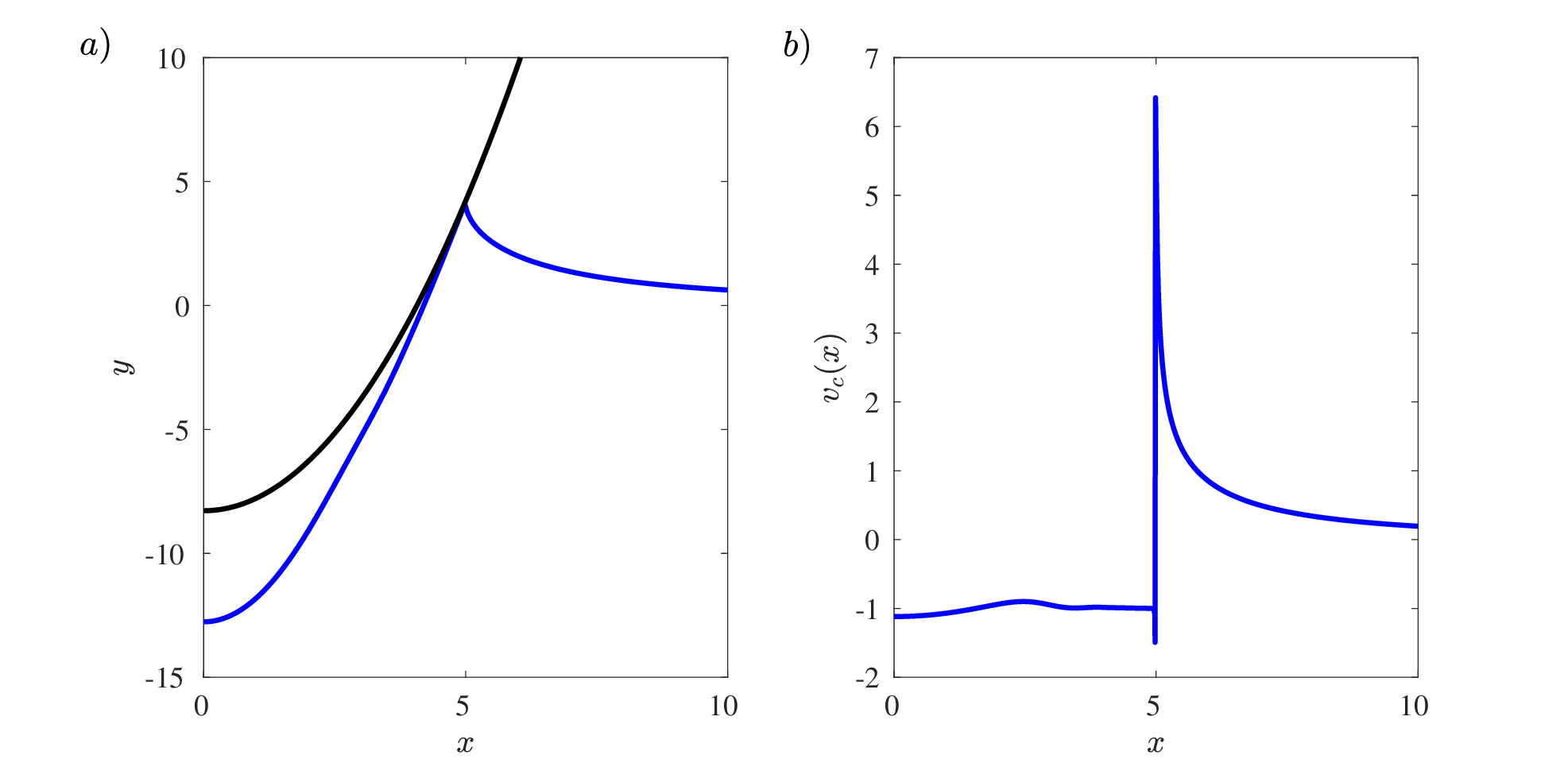}}
\caption{Numerical solution of (\ref{eqn:PreImpact1}), (\ref{eqn:PreImpact2}) for a parabolic impactor, $f(x) = x^{2}/2$. $a)$ The touchdown free surface profile, $h_{c}(x)$, indicated by the blue line. We have also included the body profile for reference in black. For this example, the touchdown time is $t_{c} \approx 8.279$ and location is $d_{c} \approx 4.985$. $b)$ The free surface velocity, $v_{c}(x)$. While there is of course some deviation, we see that the assumption (\ref{eqn:vcassumption}) is not unreasonable at touchdown, with the air approximately moving down at the same velocity as the impactor.}
\label{fig:touchdown}
\end{figure}

We show example touchdown profiles of $h_{c}(x)$ and $v_{c}(x)$ for the impact of the parabola $f(x) = x^{2}/2$ in figure \ref{fig:touchdown}. By symmetry, we just consider $x>0$. For this body, touchdown occurs at $t_{c} \approx 8.279$ with $d_{c} \approx 4.985$. In figure \ref{fig:touchdown}a, we can clearly see the entrapped air bubble at touchdown alongside the corresponding deformation of the free surface. The free surface velocity at touchdown, $v_{c}(x)$, is shown in figure \ref{fig:touchdown}b. Between the contact points, we see that, as reported in \cite{Ross2019}, the air that is trapped between the body and the liquid is moving down almost uniformly with the impact velocity, so in effect it is an extension of the solid body itself. This inspires our modelling assumption for the post-impact analysis, where we take
\begin{linenomath}
 \begin{equation}
  v_{c}(x) = -1 \quad \mbox{for} \quad -d_{c} < x < d_{c}.
 \label{eqn:vcassumption}
 \end{equation}
\end{linenomath}
going forward. In practice and as seen in figure \ref{fig:touchdown}b, this limit is not identically realized. As discussed above, the numerical pre-impact solution will need to be cut off at some point when the gap between the free surface and the solid body reaches a certain height, and so there are limitations in terms of convergence of the code. Moreover, the free surface velocity will not be identically $-1$, particularly near the peaks, where the rapid changes in free surface just before touchdown mean that $v_{c} > 0$ close to $\pm d_{c}$. However, for the majority of the interval $-d_{c}<x<d_{c}$, (\ref{eqn:vcassumption}) is a good approximation, and it allows us to readily adapt the classical Wagner post-impact model to incorporate these crucial pre-impact air interactions.

After touchdown, the air pocket pinches off, and it retracts and collapses into one or many bubbles, \cite{Marston2011, Thoroddsen2010}. For the purposes of our post-impact analysis, we will neglect the effects of this process on the fluid flow, assuming it has a lower-order effect on the resulting impact problem. For this to be a reasonable assumption, the timescale for the collapse of the bubble needs to be longer than that for the Wagner interactions that we discuss in \textsection \ref{sec:PostImpact}. A simple model for the collapse of an air bubble proposed by \cite{Marston2011, Oguz1989, Thoroddsen2005} is
\begin{linenomath}
 \begin{equation}
 \frac{\mbox{d}R^{*}_{\mathrm{bub}}}{\mbox{d}t^{*}} = -C\sqrt{\frac{\sigma}{\rho_{l}V_{\mathrm{bub}}^{*}}}R_{\mathrm{bub}}^{*}, 
 \end{equation} 
\end{linenomath}
where $R^{*}_{\mathrm{bub}}$ is the radius of the bubble, $V_{\mathrm{bub}}^{*}$ is the initial bubble volume and $C$ is a constant. Experimental analysis of solid spheres impacting onto a liquid pool in \cite{Marston2011} show that $C\propto\Oh^{-1/2}$ where $\Oh = \mu_{l}/\sqrt{\rho_{l}\sigma R^{*}}$ is the Ohnesorge number. Hence, if we approximate the volume of the entrapped bubble by $V_{\mathrm{bub}} \sim \ve^{3} R^{*3}$, the timescale for bubble collapse is given by
\begin{linenomath}
 \begin{equation}
  t_{\mathrm{bub}}^{*} = \Oh^{1/2}\sqrt{\frac{\rho_{l}\ve^{3}R^{*3}}{\sigma}}.
 \end{equation}
\end{linenomath}
On the other hand, as we shall see shortly, the Wagner timescale is the same as that for the pre-impact air-water interactions and is thus given by
\begin{linenomath}
 \begin{equation}
  t_{\mathrm{Wagner}}^{*} = \frac{\ve^{2}R^{*}}{V^{*}},
 \end{equation}
\end{linenomath}
so that our condition for neglecting the collapse of the entrapped air bubble in our analysis reduces to
\begin{linenomath}
 \begin{equation}
  t_{\mathrm{Wagner}}^{*} \ll t_{\mathrm{bub}}^{*} \implies  \ve \ll \Oh\We.
 \end{equation}
\end{linenomath}

To see whether this assumption is reasonable, let us consider an illustrative example of an impactor of typical radius of curvature $R^{*} = 1$m impacting a deep liquid pool at speed $V^{*} = 1$ms$^{-1}$. By (\ref{eqn:epsilon}), we find that $\ve \approx 2.5\times10^{-3}$, while $\Oh\We \approx 1.7$, so that the timescale for bubble collapse is much longer than that for the post-impact Wagner interactions. We shall neglect the bubble henceforth.

Finally, we shall assume that $\phi_{c}$, $h_{c}$ and $v_{c}$ all admit Taylor series at each $x$. This is certainly reasonable in terms of the velocity potential, which \cite{Ross2019} show is smooth along the liquid free surface for a range of different impactors. For $h_{c}$ and $v_{c}$, the only points that may present an issue are $\pm d_{c}$, at which the free surface becomes quite sharply peaked (cf. figure \ref{fig:touchdown}). However, for all practical applications, the code must be terminated when the minimum air gap is small but finite, so that the free surface is still smooth. Moreover, other physical effects such as surface tension will smooth out any local peaks, so again, our assumptions are reasonable.

\subsection{Key asymptotic results} \label{sec:PreImpactAsymptotics}

Before proceeding to the post-impact analysis, however, we note some important asymptotic behaviours of the solution of (\ref{eqn:PreImpact1}), (\ref{eqn:PreImpact2}), that are of relevance in both the numerical methodology and in our analysis of the post-impact behaviour in \textsection \ref{sec:PostImpact}--\ref{sec:PostImpactExample}. 

\subsubsection{`Far-away' solution, $t\rightarrow-\infty$}

In order to initiate our numerical solutions, we need to choose a time station when the impactor is far from the fluid, so that $t$ is large and negative. We shall assume that the impactor has a power-law profile $f(x) = a|x|^{n}$, where $n \geq 1$ to retain convexity of the impactor and $a>0$. 

Then, from (\ref{eqn:PreImpact1}), it is clear there are two distinct spatial regimes. When $x = O(1)$, assuming that $h$ decays at infinity --- which is physically desirable --- we find the pressure gradient is given by
\begin{linenomath}
 \begin{equation}
  Q = \frac{1}{(-t)^{3}}\left(-12x\right) + o((-t)^{-3}) \quad \mbox{as} \quad t\rightarrow-\infty.
 \end{equation}
\end{linenomath}
However, far from the line of symmetry, when $x = \hat{x}(-t)^{1/n}$, we now have
\begin{linenomath}
\begin{equation}
  Q = \frac{1}{(-t)^{3-1/n}}\left(\frac{-12\hat{x}}{(a|\hat{x}|^{n}+1)^{3}}\right) + o((-t)^{-3+1/n}) \quad \mbox{as} \quad t\rightarrow-\infty.
\end{equation}
\end{linenomath}
Thus, 
\begin{linenomath}
 \begin{equation}
  Q = \frac{-12x}{(a|x|^{n} - t)^{3}} + o\left(\frac{1}{(-t)^{3}}\right)
  \label{eqn:Qasymp}
 \end{equation}
\end{linenomath}
is a consistent leading-order solution for $Q$ valid for all $x$ as $t\rightarrow-\infty$. 

Having determined $Q$, we can find the leading-order form of the free surface by considering (\ref{eqn:PreImpact2}), yielding
\begin{linenomath}
\begin{equation}
h = \frac{6}{\pi}\dashint_{-\infty}^{\infty}\frac{s}{(a|s|^{n} - t)} \, \frac{\mbox{d}s}{x-s} + o(1) \label{eqn:Hasymp}
 \end{equation}
\end{linenomath}
as $t\rightarrow-\infty$. This is again valid for all $x$. In particular, we note that, at the line of symmetry, we have
\begin{equation}
 h(0,t) = \frac{-12}{a^{1/n}n}\mbox{cosec}\left(\frac{\pi}{n}\right)(-t)^{-1/n} + o\left((-t)^{-1/n}\right) \label{eqn:HasympLOS}
\end{equation}
as $t\rightarrow-\infty$.

The form of (\ref{eqn:Hasymp}) has important consequences for our numerical methodology. For a fixed given initial time, $-T$, it is not sufficient to simply start with no air flow and a flat free surface, particularly as $n$ gets larger. Indeed, unless we were to pick $T$ to be extremely large, the error introduced by ignoring the far-away asymptotic behaviour would dominate any introduced by a reasonable discretization of (\ref{eqn:PreImpact1}), (\ref{eqn:PreImpact2}). In fact, depending on the choice of $T$, it may be necessary to obtain the second-order term for $h$ to achieve reasonable accuracy in the numerical solution. While this analysis is fairly straightforward, we omit it here since the algebra becomes somewhat complicated and uninformative for general $n$.

\subsubsection{Large-$|x|$ solution}

When we consider the behaviour a long time after touchdown in \textsection \ref{sec:PostImpactExample}, it is necessary to know the asymptotic form of the touchdown free surface profile $h_{c}(x)$ and its velocity $v_{c}(x)$ for large $|x|$. 

Let us suppose that $t = O(1)$ and that $x\rightarrow\infty$; we shall consider positive $x$ without less of generality by the symmetry of the problem. Then, for the same power-law body considered in the previous section, it is immediate from (\ref{eqn:PreImpact1}) that the pressure gradient has the form
\begin{linenomath}
 \begin{equation}
  Q = \frac{1}{x^{3n-1}}\left(\frac{-12}{a^{3n}}\right) + o(x^{-3n+1}) \quad \mbox{as} \quad x\rightarrow\infty.
 \end{equation}
\end{linenomath}
We note that the requirement of convexity is important here, as it leads to the pressure decaying in the far-field, since $3n-1>0$.

Upon expanding the right-hand side of (\ref{eqn:PreImpact2}), we find that the contributions from the singular part of the integral are lower-order, so that
\begin{linenomath}
\begin{equation}
 \dashint_{-\infty}^{\infty} \frac{Q(\xi,t)}{\xi-x}\,\mbox{d}\xi = -\frac{1}{x}\int_{-\infty}^{\infty}Q(s,t)\,\mbox{d}s - \frac{1}{x^{2}}\int_{-\infty}^{\infty}sQ(s,t)\,\mbox{d}s + o(x^{-2}) \quad \mbox{as} \quad x\rightarrow\infty.
\end{equation}
\end{linenomath}
Therefore, after recalling that $Q(s,t)$ is an odd function so that the first term vanishes, we see that
\begin{linenomath}
\begin{equation}
 h_{c} = \frac{A(t)}{x^{2}} + o(x^{-2}), \quad v_{c} = \frac{\dot{A}(t)}{x^{2}} + o(x^{-2}) \quad \mbox{as} \quad x\rightarrow\infty,
\end{equation}
\end{linenomath}
where $A(t)$ depends on the pressure gradient across the whole of the domain. This ellipticity leads to the interesting result that the free surface decays quadratically in $x$ irrespective of the value of $n\geq1$.

%%%%%%%%%%%%%%%%%%%%%%%%%%%%%%%%%%%%%%%%%%%%%%%%%%%%%%%%%%%%%%%%%%%%%%%%%%%%
%%%%%%%%%%%%%%%%%%%%%%%%%%%%%%%%%%%%%%%%%%%%%%%%%%%%%%%%%%%%%%%%%%%%%%%%%%%%
%%%%%%%%%%%%%%%%%%%%%%%%%%%%%%%% POST-IMPACT %%%%%%%%%%%%%%%%%%%%%%%%%%%%%%%
%%%%%%%%%%%%%%%%%%%%%%%%%%%%%%%%%%%%%%%%%%%%%%%%%%%%%%%%%%%%%%%%%%%%%%%%%%%%
%%%%%%%%%%%%%%%%%%%%%%%%%%%%%%%%%%%%%%%%%%%%%%%%%%%%%%%%%%%%%%%%%%%%%%%%%%%%

\section{Post-impact model: Wagner theory} \label{sec:PostImpact}

\begin{figure}[t!]
\centering \scalebox{0.7}{\epsfig{file=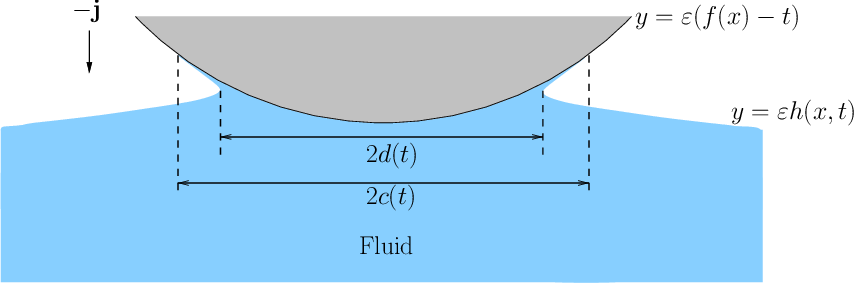}}
\caption{The post-impact configuration. Both the entrapped air bubble and the air ahead of the turnover points, $x = \pm d(t)$ are neglected. The liquid free surface is violently disturbed by the impacting body, forcing liquid into two splash jets that are attached to the impactor. The wetted extent of the body, between the tips of these jets, is $2c(t)$.}
\label{fig:PostImpactStructure} 
\end{figure}

At $t = t_{c}$, since we ignore the role of the entrapped gas bubble, the solid body is assumed to be wetted over the region $|x|<d_{c}$, while the remainder of the body is not in contact with the liquid. As time progresses further, the free surface is violently displaced into two high-speed, thin jets of liquid that shoot along the impacting body. For $t>t_{c}$, the picture is as in figure \ref{fig:PostImpactStructure}, in which $x = \pm d(t)$ now denote the turnover points where the free surface becomes vertical and $x = \pm c(t)$ denote the tips of the jets (hence the extent of wetted region). For the remainder of this analysis, we shall assume that the air layers between the jets/impactor and the liquid pool play a lower-order role in the evolution of the impact. Hence, in particular, the jets are assumed to remain attached to the solid body and the air between the body and ahead of the wetted region is essentially quiescent. The reader is directed to \cite{Purvis2004, Moore2013, Moore2014} for a detailed investigation into the effect of these air regions after impact.

To investigate the post-impact behaviour, we shall follow the ideas of Wagner theory, \cite{Wagner1932, Howison1991}. On the lengthscale $O(\ve R^{*})$, which we recall was the appropriate lengthscale in the liquid for the pre-impact dynamics discussed in \textsection \ref{sec:PreImpact}, the boundary conditions linearize onto the undisturbed waterline, and the effect of the jets can be neglected. Then, for $t>t_{c}$, to leading order, the liquid velocity potential must therefore satisfy the mixed boundary value problem
\refstepcounter{equation}
\begin{linenomath}
$$
 \nabla^{2}\phi = 0, \quad \frac{\partial\phi}{\partial t} = -p \quad \mbox{in} \quad y <0,
 \eqno{(\theequation{\mathit{a},\mathit{b}})}
 \label{eqn:PostImpactOuterEqns}
$$
\end{linenomath}
such that
\begin{linenomath}
\begin{equation}
 \frac{\partial\phi}{\partial y} = -1 \quad \mbox{on} \quad y = 0, |x|<d(t)
 \label{eqn:PostImpactBodyBC5}
\end{equation}
\end{linenomath}
on the solid body, and 
\refstepcounter{equation}
\begin{linenomath}
$$
 p = 0, \quad \frac{\partial \phi}{\partial y} = \frac{\partial h}{\partial t}, \quad \mbox{on} \quad y = 0, |x|>d(t) 
  \eqno{(\theequation{\mathit{a},\mathit{b}})}
 \label{eqn:PostImpactInterfaceBCs}
$$
\end{linenomath}
on the free surface. Far from the impact, the fluid is still quiescent, so that
\begin{linenomath}
\begin{equation}
 \nabla\phi \rightarrow 0 \quad \mbox{as} \quad x^{2}+y^{2}\rightarrow\infty, \quad h \rightarrow0 \quad \mbox{as} \quad |x|\rightarrow\infty,
 \label{eqn:PostImpactFFC}
\end{equation}
\end{linenomath}
while the initial conditions at $t = t_{c}$ are given by (\ref{eqn:InitialConditions}) and $d(t_{c}) = d_{c}$.

\begin{figure}[t!]
\centering \scalebox{0.9}{\epsfig{file=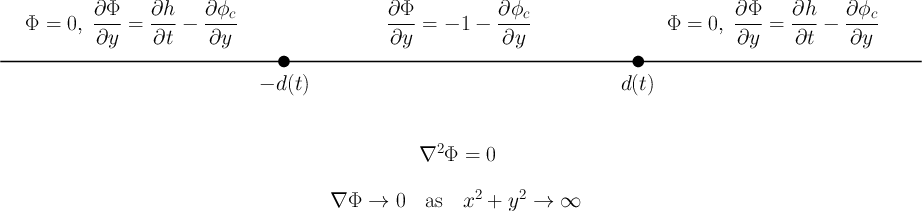}}
\caption{The leading-order mixed boundary value problem for the $\Phi$. A local analysis reveals that $\Phi \sim S(t)\sqrt{(x+\mbox{i}y)^{2}-d(t)^{2}})$ as $|x+\mbox{i}y \mp d(t)|\rightarrow0$.}
\label{fig:PostImpactOuterMBVP} 
\end{figure}

By using the linearized Bernoulli equation (\ref{eqn:PostImpactOuterEqns}b), we can integrate the dynamic boundary condition (\ref{eqn:PostImpactInterfaceBCs}a), to deduce that $\phi = \phi_{c}(x,0)$ on $y = 0$, $|x|>d(t)$. Then, writing $\Phi = -\phi_{c}(x,y) + \phi$, we transform (\ref{eqn:PostImpactOuterEqns})--(\ref{eqn:PostImpactFFC}) to the mixed boundary value problem depicted in figure \ref{fig:PostImpactOuterMBVP}. In particular, we note that the Laplace equation is unchanged since $\phi_{c}(x,y)$ must be harmonic in the lower half-plane. Moreover, the far-field condition is also unchanged by this transformation. 

The problem in figure \ref{fig:PostImpactOuterMBVP} is codimension two in the sense that $d(t)$ must be determined as part of the solution, \cite{Howison1997}. The region $-d(t)<x<d(t)$ is often referred to as the \emph{effective contact set} in impact literature, as it is the pressure in this region that provides the dominant contribution to the slamming force. In particular, the change in boundary condition type from the effective contact set $|x|<d(t)$ to the non-contact set $|x|>d(t)$ drives a singularity in the flow. Since $\phi_{c}(x,0)$ is assumed to have a valid Taylor expansion as $x\rightarrow\pm d(t)$, it is readily shown that the local analysis in \cite{Gillow1998} can be extended to the present case and that the least singular solution must be such that
\begin{linenomath}
\begin{equation}
 \Phi = O\left(\sqrt{z^{2}-d(t)^{2}}\right) \quad \mbox{as} \quad |z\mp d(t)|\rightarrow0,
\end{equation}
\end{linenomath}
where we have introduced the complex number $z = x+\mbox{i}y$ for convenience.

In essence therefore, we have transformed the leading-order post-impact problem to the traditional Wagner `outer' flat-plate problem \cite{Howison1991, Wagner1932}, with an appropriate choice of impact velocity given by
\begin{linenomath}
\begin{equation}
\mathcal{V} = -1 - \frac{\partial\phi_{c}}{\partial y}(x,0) = -1 - v_{c}(x). 
\end{equation}
\end{linenomath}
By considering the function 
\begin{linenomath}
 \begin{equation}
  \mathcal{G} = \mbox{i}\sqrt{z^{2}-d(t)^{2}}\frac{\mbox{d}W}{\mbox{d}z},
 \end{equation}
\end{linenomath}
where $W = \Phi + \mbox{i}\Psi$ is the complex potential and $\Psi(x,y,t)$ the streamfunction, we transform the problem in figure \ref{fig:PostImpactOuterMBVP} to a Dirichlet problem for $\mbox{Im}(\mathcal{G})$. The branch cut for the square root function is taken along the real axis between the turnover points, and is such that the function is real and positive for $x>d(t)$. After solving the Dirichlet problem, we find that the complex velocity is thus given by
\begin{linenomath}
 \begin{equation}
  \frac{\mbox{d}W}{\mbox{d}z} = \mbox{i}\left(1-\frac{z}{\sqrt{z^{2}-d(t)^{2}}}\right) + \frac{\mbox{i}}{\pi\sqrt{z^{2}-d(t)^{2}}}\int_{-d(t)}^{d(t)}\frac{\sqrt{d(t)^{2}-s^{2}}}{s-z}v_{c}(s)\,\mbox{d}s.
  \label{eqn:LOOComplexVelocity}
 \end{equation}
\end{linenomath}
In (\ref{eqn:LOOComplexVelocity}), the first term is simply the classical Wagner solution for the constant-speed impact of a rigid body with no air cushioning, while the integral term is the correction due to the pre-impact behaviour.

Applying the kinematic condition on the free surface (\ref{eqn:PostImpactInterfaceBCs}b) and integrating with respect to time, we find that the free surface profile is given by
\begin{linenomath}
 \begin{alignat}{2}
  h & = &&\, -\left(1 - v_{c}\right)(t-t_{c}) + \int_{t_{c}}^{t} \frac{x}{\sqrt{x^{2}-d(\tau)^{2}}}\,\mbox{d}\tau + h_{c}(x) - \frac{1}{\pi}\int_{t_{c}}^{t}\frac{1}{\sqrt{x^{2}-d(\tau)^{2}}}\int_{-d(\tau)}^{d(\tau)} \frac{\sqrt{d(\tau)^{2}-s^{2}}}{s-x}v_{c}(s)\,\mbox{d}s\,\mbox{d}\tau
  \label{eqn:OuterFreeSurface}
 \end{alignat}
\end{linenomath}
for $x>d(t)$, and we can recover $h$ for $x<-d(t)$ by symmetry.

It remains to determine the evolution of the turnover point, $d(t)$. Since the velocity and pressure are both singular at $x = \pm d(t)$ (cf. (\ref{eqn:LOOComplexVelocity})), the `outer' solution breaks down, and we need to consider an `inner' region in which the free surface turns over and fluid is forced into the high-speed, thin liquid jet. As discussed in detail in Appendix \ref{sec:AppendixWagnerInner}, these local regions centred on $x = \pm d(t), y = \ve(f(d(t))-t)$ have a typical lengthscale that is two orders of magnitude smaller than that in the outer region, so that $x\mp d(t) = O(\ve^{2})$ and $y -\ve(f(d(t))-t)= O(\ve^{2})$. In this region, the leading-order problem is a quasi-steady Helmholtz flow and, in particular, matching between the leading-order-outer and leading-order-inner free surface profiles necessitates that the so-called \emph{Wagner condition} must hold:
\begin{linenomath}
 \begin{equation}
  h(d(t),t) = f(d(t))-t \quad \mbox{for} \quad t>t_{c}.
  \label{eqn:WagnerCondition}
 \end{equation}
\end{linenomath}
In essence, this is simply a statement of conservation of mass, in that it enforces the outer free surface profile to meet the impacting body at the turnover point in this outer region, \cite{Wagner1932, Howison1991}.

By performing an asymptotic expansion of (\ref{eqn:OuterFreeSurface}) close to the turnover point $x = d(t)$, we find that
\begin{linenomath}
\begin{alignat}{2}
 h & \,= &&\, h_{c}(d(t)) -\left(1 - v_{c}(d(t))\right)(t-t_{c}) + \int_{t_{c}}^{t}\frac{d(t)}{\sqrt{d(t)^{2}-d(\tau)^{2}}}\,\mbox{d}\tau \nonumber \\
 & && \,- \frac{1}{\pi}\int_{t_{c}}^{t}\frac{d(\tau)^{2}}{\sqrt{d(t)^{2}-d(\tau)^{2}}}\int_{-1}^{1} \frac{\sqrt{1-S^{2}}}{d(\tau)S-d(t)}v_{c}(d(\tau)S)\,\mbox{d}S\,\mbox{d}\tau \nonumber \\
 & \, && \, +\left[\frac{\sqrt{2d(t)}}{\dot{d}(t)}\left(1+\frac{1}{\pi}\int_{-1}^{1}\sqrt{\frac{1+S}{1-S}} v_{c}(d(t)S)\right) \,\mbox{d}S\right]\sqrt{x-d(t)} + o\left(\sqrt{x-d(t)}\right)
\end{alignat}
\end{linenomath}
as $x-d(t)\rightarrow0^{+}$. Hence, the Wagner condition (\ref{eqn:WagnerCondition}) gives
\begin{linenomath}
 \begin{alignat}{2}
  f(d(t))-t & =  &&\, h_{c}(d(t)) -\left(1 - v_{c}(d(t))\right)(t-t_{c}) + \int_{t_{c}}^{t}\frac{d(t)}{\sqrt{d(t)^{2}-d(\tau)^{2}}}\,\mbox{d}\tau \nonumber \\
  & && \,- \frac{1}{\pi}\int_{t_{c}}^{t}\frac{d(\tau)^{2}}{\sqrt{d(t)^{2}-d(\tau)^{2}}}\int_{-1}^{1} \frac{\sqrt{1-S^{2}}}{d(\tau)S-d(t)}v_{c}(d(\tau)S)\,\mbox{d}S\,\mbox{d}\tau.
  \label{eqn:d}
 \end{alignat}
\end{linenomath}
Typically, this must be inverted numerically, although we shall discuss some asymptotic sub-limits shortly.

\subsection{Force on the impactor}

The force per unit length on the impactor post-impact, $F^{*}$, is dominated by the contribution from this outer region in which $(x^{*},y^{*}) = O(\ve R^{*})$ and $p^{*}_{l} = O(\rho_{l}^{*}V^{*2}/\ve)$, so that $F^{*}$ is $O(\rho_{l}^{*}R^{*}V^{*2})$. Even though the pressure in the inner region is an order of magnitude larger than the pressure in the outer (see Appendix \ref{sec:AppendixWagnerInner}) since the jet-root region is two orders of magnitude smaller than the outer, the force contributions are lower. The contribution from the jet itself is even smaller, as described in detail in \cite{Howison1991}. By expanding the force per unit length on the impactor in an asymptotic series of the form
\begin{linenomath}
 \begin{equation}
  F^{*} = \rho_{l}^{*}R^{*}V^{*2}\left(F_{0} + O(\ve)\right),
 \end{equation}
\end{linenomath}
we can utilize the linearized Bernoulli equation (\ref{eqn:PostImpactOuterEqns}b) to deduce that
\begin{linenomath}
 \begin{equation}
  F_{0} = -\int_{-d}^{d}\frac{\partial}{\partial t}\phi(x,0,t)\,\mbox{d}x = -\int_{-d}^{d}\frac{\partial}{\partial t}\Phi(x,0,t)\,\mbox{d}x = -\frac{\mbox{d}}{\mbox{d}t}\int_{-d}^{d}\Phi(x,0,t)\,\mbox{d}x
  \label{eqn:LOForce}
 \end{equation}
\end{linenomath}
where the final equality follows from Leibniz's rule and the fact that the velocity potential is bounded at the turnover points, $x = \pm d(t)$.

Now, we can evaluate (\ref{eqn:LOOComplexVelocity}) on $x-\mbox{i}0$, $|x|<d(t)$ and take real parts to isolate $\Phi_{x}$. Then, after integrating with respect to $x$ and recalling that $\Phi_{x}$ is square-root bounded at $x = -d(t)$, we see that
\begin{linenomath}
\begin{equation}
 \Phi(x,0,t) = -\sqrt{d(t)^{2}-x^{2}} - \frac{1}{\pi}\int_{-d(t)}^{x}\frac{1}{\sqrt{d(t)^{2}-\xi^{2}}}\dashint_{-d(t)}^{d(t)}\frac{\sqrt{d(t)^{2}-s^{2}}}{s-\xi}v_{c}(s)\,\mbox{d}s\mbox{d}\xi
\end{equation}
\end{linenomath}
for $|x|<d(t)$. Therefore, substituting into the leading-order force expression (\ref{eqn:LOForce}), we have
\begin{linenomath}
 \begin{equation}
  F_{0} = \pi d(t)\dot{d}(t) + \frac{1}{\pi}\frac{\mbox{d}}{\mbox{d}t}\int_{-d(t)}^{d(t)}\int_{-d(t)}^{x}\frac{1}{\sqrt{d(t)^{2}-\xi^{2}}}\dashint_{-d(t)}^{d(t)}\frac{\sqrt{d(t)^{2}-s^{2}}}{s-\xi}v_{c}(s)\,\mbox{d}s\mbox{d}\xi\mbox{d}x. 
 \end{equation}
\end{linenomath}
The first term is the leading-order force predicted by traditional Wagner theory in the absence of air cushioning, \cite{Howison1991}, while the second term is the correction due to the pre-impact dynamics. We can simplify this by first switching the order of the outer two integrals, and integrating the result with respect to $x$, yielding
\begin{linenomath}
 \begin{equation}
  F_{0} = \pi d(t)\dot{d}(t) + \frac{1}{\pi}\frac{\mbox{d}}{\mbox{d}t}\int_{-d(t)}^{d(t)}\sqrt{\frac{d(t)-\xi}{d(t)+\xi}}\dashint_{-d(t)}^{d(t)}\frac{\sqrt{d(t)^{2}-s^{2}}}{s-\xi}v_{c}(s)\,\mbox{d}s\mbox{d}\xi. 
 \end{equation}
\end{linenomath}
Finally, we can switch the order of integration again and evaluate the principal value integral, leaving us with the expression
\begin{linenomath}
 \begin{equation}
  F_{0} = \pi d(t)\dot{d}(t) \left(1 + \frac{1}{\pi}\int_{-d(t)}^{d(t)}\frac{1}{\sqrt{d(t)^{2}-s^{2}}}v_{c}(s)\,\mbox{d}s\right)
  \label{eqn:Force}
 \end{equation}
\end{linenomath}
for the leading-order force.

It is worth noting that, since the free surface velocity just prior to touchdown is negative between the pressure peaks (see figure \ref{fig:touchdown}), we expect the air effects to reduce the force per unit length on the impactor from the classical Wagner prediction, at least initially. It is in this sense that the air `cushions' the impact. Note that, in particular, since we assume $v_{c} = -1$ between the free surface maxima at touchdown, at $t  = t_{c}$, the term in brackets in (\ref{eqn:Force}) vanishes, so that $F_{0}(t_{c}) = 0$. In practice, at touchdown the air bubble between the solid and the liquid covers the whole of the region between $\pm d_{c}$, so that we would need to consider the air pressure contributions to the slamming force.

\subsection{Outer pressure}

Finally, both for completeness and for calculating composite pressure readings on different parts of the impactor in \textsection \ref{sec:PostImpactExample}, we state that the leading-order outer pressure on the impactor is
\begin{linenomath}
\begin{equation}
 p(x,t) = \frac{d(t)\dot{d}(t)}{\sqrt{d(t)^{2}-x^{2}}} - \frac{1}{\pi}\frac{\mbox{d}}{\mbox{d}t}\int_{-d(t)}^{x}\frac{1}{\sqrt{d(t)^{2}-\xi^{2}}}\dashint_{-d(t)}^{d(t)}\frac{\sqrt{d(t)^{2}-s^{2}}}{s-\xi}v_{c}(s)\,\mbox{d}s\mbox{d}\xi
 \label{eqn:OuterPressure}
\end{equation}
\end{linenomath}
for $|x|<d(t)$. To construct the composite pressure profile, we will need the local expansion of the pressure close to the turnover point, which is given by
\begin{linenomath}
\begin{equation}
 p(x,t) \sim \frac{\dot{d}(t)\sqrt{d(t)}}{\sqrt{2}}\left(1+\frac{1}{\pi d(t)}\int_{-d(t)}^{d(t)}\sqrt{\frac{d(t)+s}{d(t)-s}}v_{c}(s)\,\mbox{d}s\right)\frac{1}{\sqrt{d(t)-x}}
 \label{eqn:OuterPressureExpansion}
\end{equation}
\end{linenomath}
as $d(t)-x\rightarrow0^{+}$. Note, again, for times just after touchdown, since $v_{c}(x)$ is predominantly negative for $-d(t)<x<d(t)$, the second term in (\ref{eqn:OuterPressureExpansion}) reduces the coefficient of the square-root singularity from the classical Wagner solution, which is simply the first term in (\ref{eqn:OuterPressure}).

\subsection{Jet thickness}

While we relegate a detailed discussion of the inner region local to the jet-roots to Appendix \ref{sec:AppendixWagnerInner}, an important prediction from the local analysis is the asymptotic jet thickness at its base, denoted by $J(t)$. Fluid is ejected into the splash jets from the inner region at at speed $U(t) = 2\dot{d}(t)$, so that $U(t)J(t)$ gives the leading-order dimensionless mass of fluid ejected in the splash. In Appendix \ref{sec:AppendixWagnerInner}, we show that
\begin{linenomath}
 \begin{equation}
  J(t) = \frac{\pi d(t)}{8\dot{d}(t)^{2}}\left(1+\frac{1}{\pi}\int_{-1}^{1}\sqrt{\frac{1+S}{1-S}}v_{c}(d(t)S)\,\mbox{d}S\right)^{2}.
  \label{eqn:JetThickness}
 \end{equation}
\end{linenomath}
As with the slamming force, by the assumption that $v_{c} = -1$ for $-d_{c}<x<d_{c}$, at $t = 0$, we have $J(t) = 0$. Moreover, since the integral is negative, at least for small times, the jet thickness is reduced from the classical Wagner solution, which is simply given by the first term in the brackets on the right-hand side of (\ref{eqn:JetThickness}) (see, for example, \cite{Howison1991}).

\section{Post-impact results for a power-law body profile} \label{sec:PostImpactExample}

Let us consider the post-impact behaviour for an impactor that has a power-law profile of the form
\begin{linenomath}
\begin{equation}
 f(x)  = a|x|^{n}
\end{equation}
\end{linenomath}
where $n\geq 1$ and $a > 0$. In the absence of air cushioning, Wagner theory gives the leading-order location of the turnover point to be
\begin{linenomath}
\begin{equation}
 d_{\mathrm{Wagner}}(t) = \left(\frac{\pi}{2^{n}aB((n+1)/2,(n+1)/2)}\right)^{1/n}t^{1/n},
\label{eqn:Wagnerd}
 \end{equation}
\end{linenomath}
where $B(\cdot,\cdot)$ is the Beta function, and the leading-order force on the impactor is
\begin{linenomath}
 \begin{equation}
 F_{\mathrm{Wagner}} = \pi d_{\mathrm{Wagner}}(t)\dot{d}_{\mathrm{Wagner}}(t),
 \label{eqn:WagnerF}
 \end{equation}
\end{linenomath}
as shown by \cite{Moore2014b}. In this section, we will investigate the influence of pre-impact cushioning on these results, and, in particular, consider the dynamics just after and a long time after touchdown.

\subsection{Small-time solution} \label{sec:SmallTimeSoln}

We shall begin by considering the behaviour just after touchdown, so that $t - t_{c} \rightarrow0$. We seek a solution for the turnover point location by considering the asymptotic expansion
\begin{linenomath}
 \begin{equation}
  d(t) = d_{c} + \tilde{d}_{1}(t-t_{c})^{\beta} + o((t-t_{c})^{\beta}) \quad \mbox{as} \quad t - t_{c} \rightarrow0,
 \end{equation}
\end{linenomath}
where $\beta, \tilde{d}_{1}> 0$. Upon substituting this expansion into the Wagner condition (\ref{eqn:d}) and recalling that, by definition of touchdown, $ad_{c}^{n} - t_{c} - h_{c}(d_{c}) = 0$, we see that
\begin{linenomath}
\begin{alignat}{2}
& && \,\left(nad_{c}^{n-1} - h_{c}'(d_{c})\right)\tilde{d}_{1}(t - t_{c})^{\beta} - v_{c}(d_{c})(t - t_{c}) + o((t - t_{c})) \nonumber \\
& && \, \qquad \qquad = \sqrt{\frac{2d_{c}}{\tilde{d}_{1}\beta}}\left(1+\frac{1}{\pi}\int_{-1}^{1}\sqrt{\frac{1+s}{1-s}}v_{c}(d_{c}s)\,\mbox{d}s\right)(t - t_{c})^{1-\beta/2} + o((t - t_{c})^{1-\beta/2})
\end{alignat}
\end{linenomath}
as $t - t_{c}\rightarrow0$, where a prime indicates differentiation with respect to argument. Clearly, to retain a dominant balance, we must take $\beta = 2/3$, with the coefficient of the correction to the location of the turnover point then given by
\begin{linenomath}
\begin{equation}
 \tilde{d}_{1} = \left[\frac{3\sqrt{d_{c}}\pi}{4\sqrt{2}(nad_{c}^{n-1}-h_{c}'(d_{c}))}\left(1+\frac{1}{\pi}\int_{-1}^{1}\sqrt{\frac{1+s}{1-s}}v_{c}(d_{c}s)\,\mbox{d}s\right)\right]^{2/3}.
 \label{eqn:SmallTimed}
\end{equation}
\end{linenomath}
In the interests of considering the force on the impactor, it is relatively straightforward to show that the correct form for the next term in the asymptotic expansion for the turnover point location is $\tilde{d}_{2}(t-t_{c})^{4/3}$, but given the algebraic complication, we do not solve for $\tilde{d}_{2}$ here.

Clearly, this behaviour is somewhat different to the classical Wagner solution (\ref{eqn:Wagnerd}). Naturally, this is due to the fact that the initial contact set has non-zero measure, and indeed, the $2/3$-scaling also occurs in the water entry of a flat-bottomed wedge in the absence of air cushioning, \cite{Oliver2002}. 

We can use this knowledge of the early-time turnover point behaviour to investigate the leading-order force on the impactor. Expanding (\ref{eqn:Force}) as $t-t_{c}\rightarrow0$, we find that
\begin{linenomath}
 \begin{equation}
  F(t) = \tilde{F}_{0}(t-t_{c})^{-1/3} + \tilde{F}_{1}(t-t_{c})^{1/3} + o\left((t-t_{c})^{1/3}\right)
 \end{equation}
\end{linenomath}
where
\refstepcounter{equation}
\begin{linenomath}
$$
  \tilde{F}_{0} = \frac{2\pi d_{c}\alpha}{3}\left(1+\frac{1}{\pi}\int_{-1}^{1}\frac{v_{c}(d_{c}s)}{\sqrt{1-s^{2}}}\,\mbox{d}s\right), \quad \tilde{F}_{1} = \frac{\tilde{F}_{0}}{\alpha d_{c}}\left(\alpha^{2} + 2d_{c}d_{2}\right) +  \frac{2\pi d_{c}\alpha^{2}}{3}\int_{-1}^{1}\frac{sv_{c}'(d_{c}s)}{\sqrt{1-s^{2}}}\,\mbox{d}s.
\eqno{(\theequation{\mathit{a},\mathit{b}})}
\label{eqn:SmallTimeForce}
$$
\end{linenomath}
We have included both of these terms to stress an important point when using numerical data of the pre-impact behaviour. In the context of the model, in which we have assumed that (\ref{eqn:vcassumption}) holds, the integral in (\ref{eqn:SmallTimeForce}a) is identically $-\pi$ so that $\tilde{F}_{0}$, and hence the force on the impactor is bounded as we approach touchdown. This is expected in the post-impact model, which essentially does not see the air bubble at all. However, in practice, when using simulation data, (\ref{eqn:vcassumption}) will only be approximately true, so that when using numerical data, we would expect the predicted post-impact force to be singular just after touchdown. 

This is certainly a drawback of using the new model with real numerical data, but it is not a major concern. The model is not valid for small scales just after touchdown in general, since the air bubble would still be expected to cover a significant portion of the impactor between $\pm d(t)$. We expect that there is a transition from the viscous air-cushioning forces on the impactor just before touchdown to the Wagner force just after touchdown, in which the presence of the entrapped air makes an important contribution. We also note that this drawback is also true when using classical Wagner theory itself; at small times just after touchdown, there will always be discrepancies between these inviscid models and real-world data. While our current model with its appreciation of the air cushioning before touchdown shares these problems at $t-t_{c} = 0^{+}$, it is a more realistic description of the flow for later times.

Here we also note that in his model of the impact of a flat-bottomed wedge, \cite{Oliver2002} finds contributions to the leading-order force from both the inner and outer Wagner regions, rather than just the outer contribution here. This is due to the fact that the pre-impact flow is neglected in \cite{Oliver2002}, a somewhat unrealistic assumption, particularly as air-cushioning effects are likely to be most crucial for body shapes of zero deadrise angle, \cite{Wilson1991}. In \cite{Oliver2002}, the velocity potential expansion includes a Heaviside function in time to account for this sudden change after impact, which produces a leading-order force on the body comparable to that contributed by the inner region. Here, as we include the effects of the pre-impact flow and free surface deformation, the contribution to the leading-order force is dominated by the outer region, as is consistent with classical Wagner theory, \cite{Howison1991}.

\subsection{Large-time behaviour} \label{sec:LargeTimeSoln}

We now consider the opposite extreme, a long time after touchdown as $t\rightarrow\infty$. We assume that $d(t) \sim \mathring{d}_{1} t^{\gamma}$ as $t\rightarrow\infty$, where $\mathring{d}_{1} > 0$. Returning to the Wagner condition (\ref{eqn:d}), the dominant term on the left-hand side is simply the $f(d(t))$ term, which is of $O(t^{\gamma n})$ as $t\rightarrow\infty$. On the right-hand side, some careful analysis of the integrals reveals that
\begin{linenomath}
 \begin{equation}
   \int_{t_{c}}^{t}\frac{d(t)}{\sqrt{d(t)^{2}-d(\tau)^{2}}}\,\mbox{d}\tau = O(t), \quad \int_{t_{c}}^{t}\frac{d(\tau)^{2}}{\sqrt{d(t)^{2}-d(\tau)^{2}}}\int_{-1}^{1} \frac{\sqrt{1-S^{2}}}{d(\tau)S-d(t)}v_{c}(d(\tau)S,0)\,\mbox{d}S\,\mbox{d}\tau = O(t^{1-\gamma})
 \end{equation}
\end{linenomath}
as $t\rightarrow\infty$. Hence, the dominant balance enforces that $\gamma = 1/n$, and we retrieve the classical Wagner solution in the absence of air at leading order with, 
\begin{linenomath}
\begin{equation}
 \mathring{d}_{1} = \left(\frac{\pi}{2^{n}aB((n+1)/2,(n+1)/2)}\right)^{1/n},
\end{equation}
\end{linenomath}
as in (\ref{eqn:Wagnerd}).

Therefore, for large times, the effect of the air cushioning is lower-order, and the dominant behaviour is again due to the inertia of the liquid alone. This is consistent with the comparisons of Wagner theory in the absence of air cushioning with direct numerical simulations in, for example, \cite{Cimpeanu2018, Philippi2016}, where the Wagner prediction of the turnover point location is shown to be remarkably accurate once the effect of the entrapped air is accounted for in the numerical results.

The large-time solution for the force behaves similarly. It is straightforward to show that 
\begin{linenomath}
 \begin{equation}
  F = \frac{\pi \mathring{d}_{1}^{2}}{n}t^{2/n-1} + O(t^{1/n-1}) \quad \mbox{as} \quad t\rightarrow\infty,
 \end{equation}
\end{linenomath}
which is precisely that predicted by classical Wagner theory in (\ref{eqn:WagnerF}). While the air-cushioning effects come in at $O(t^{1/n-1})$, it is clear that sufficiently long after the initial touchdown, Wagner theory in the absence of air-cushioning provides a reasonable estimate of the leading-order force on the impactor.

\subsection{Results for a parabolic impactor when $t = O(1)$}

When $t = O(1)$, we in general must proceed numerically. For this reason, we shall focus our investigation on the impact of the parabola for which $f(x) = x^{2}/2$. We recall that, as discussed in \textsection \ref{sec:PreImpact}, for this body profile, $t_{c} \approx 8.279$ and $d_{c} \approx 4.985$, and $h_{c}(x)$ and $v_{c}(x)$ are depicted in figure \ref{fig:touchdown}.

\begin{figure}[t!]
\centering \scalebox{0.5}{\epsfig{file=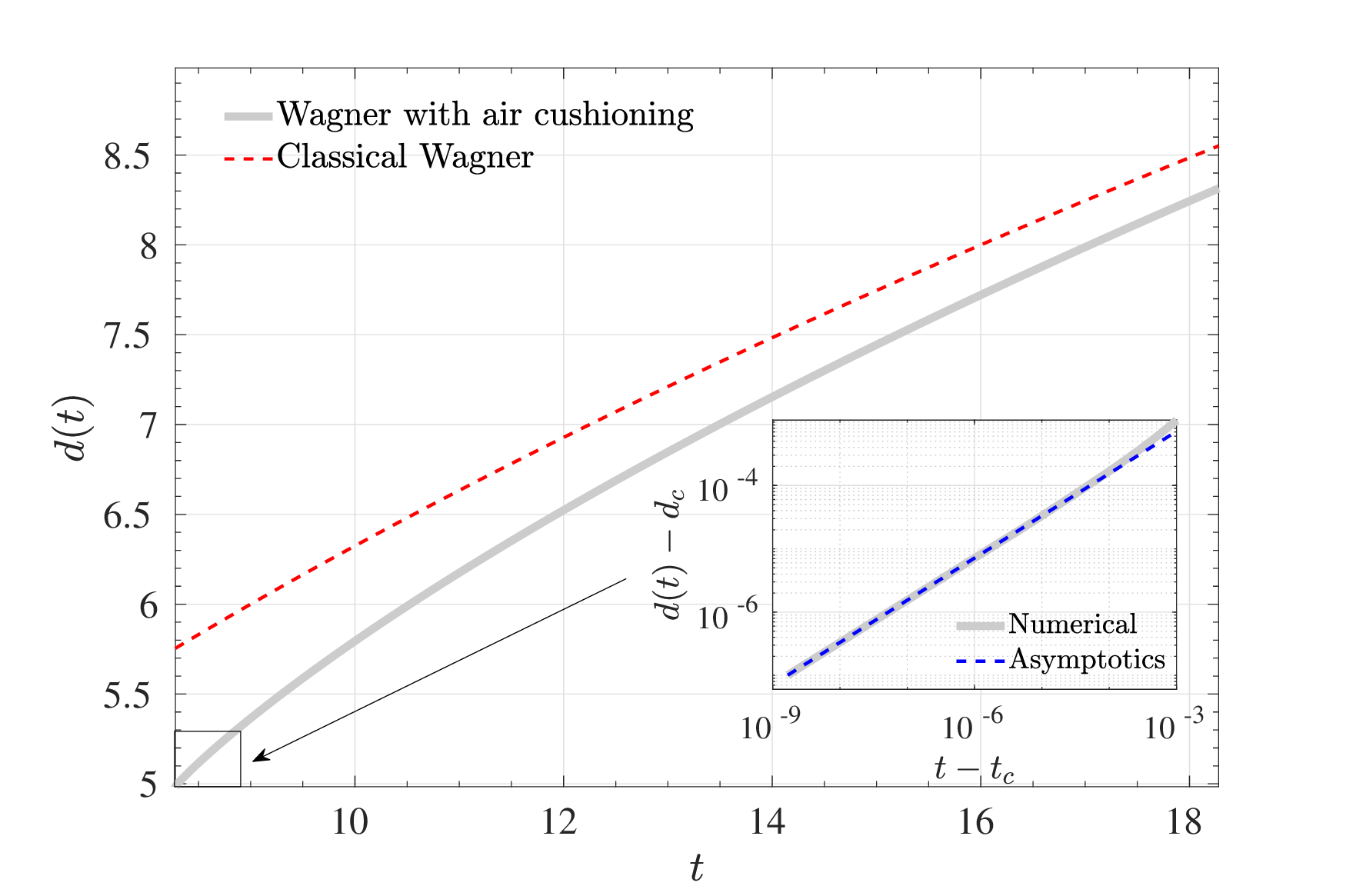}}
\caption{The leading-order turnover point location, $x = d(t)$, for the impact of the parabola $f(x) = x^{2}/2$ (solid line). In this plot, touchdown occurs at $d_{c} \approx 4.985$ at time $t_{c} \approx 8.279$. The red dashed curve is the equivalent turnover point location in the absence of air cushioning, that is $d(t) = 2\sqrt{t}$. Note that as time increases, we converge (slowly) onto this line. The inset displays a log-log plot of the behaviour just after touchdown, with the blue dashed line indicating the asymptotic limit (\ref{eqn:SmallTimed}).}
\label{fig:TurnoverPoint}
\end{figure}

Given the presence of the air-cushioning terms, inverting the integral equation (\ref{eqn:d}) is non-trivial, and must in general be performed numerically, and we describe our methodology in detail in Appendix \ref{sec:AppendixNumericalTurnoverInversion}. The results for the parabola are presented as the solid grey line in figure \ref{fig:TurnoverPoint} alongside the corresponding solution from classical Wagner theory in the absence of air cushioning, for which $t_{c} = d_{c} = 0$ and $d(t) = 2\sqrt{t}$, which is depicted as the red dashed line. It is clear that the pre-impact dynamics inhibit the growth of $d(t)$: as the body enters the liquid, the air has cushioned the impact, delaying the growth of the effective contact set. As the impact progresses, however, the effect of the air and the pre-impact dynamics fades, and we slowly converge on the Wagner solution as we found in \textsection \ref{sec:LargeTimeSoln}. In the inset of the figure, we also see the behaviour just after touchdown, where there is excellent agreement between the asymptotic solution (\ref{eqn:SmallTimed}) and the numerical solution over several decades. We note that, although their results only go up to touchdown, these results compare favourably with the results in figure 4 of \cite{Ross2019}, in which the Wagner solution over-predicts the location of the turnover curve in neglecting the effect of the air before touchdown. 

\begin{figure}[t!]
\centering \scalebox{0.5}{\epsfig{file=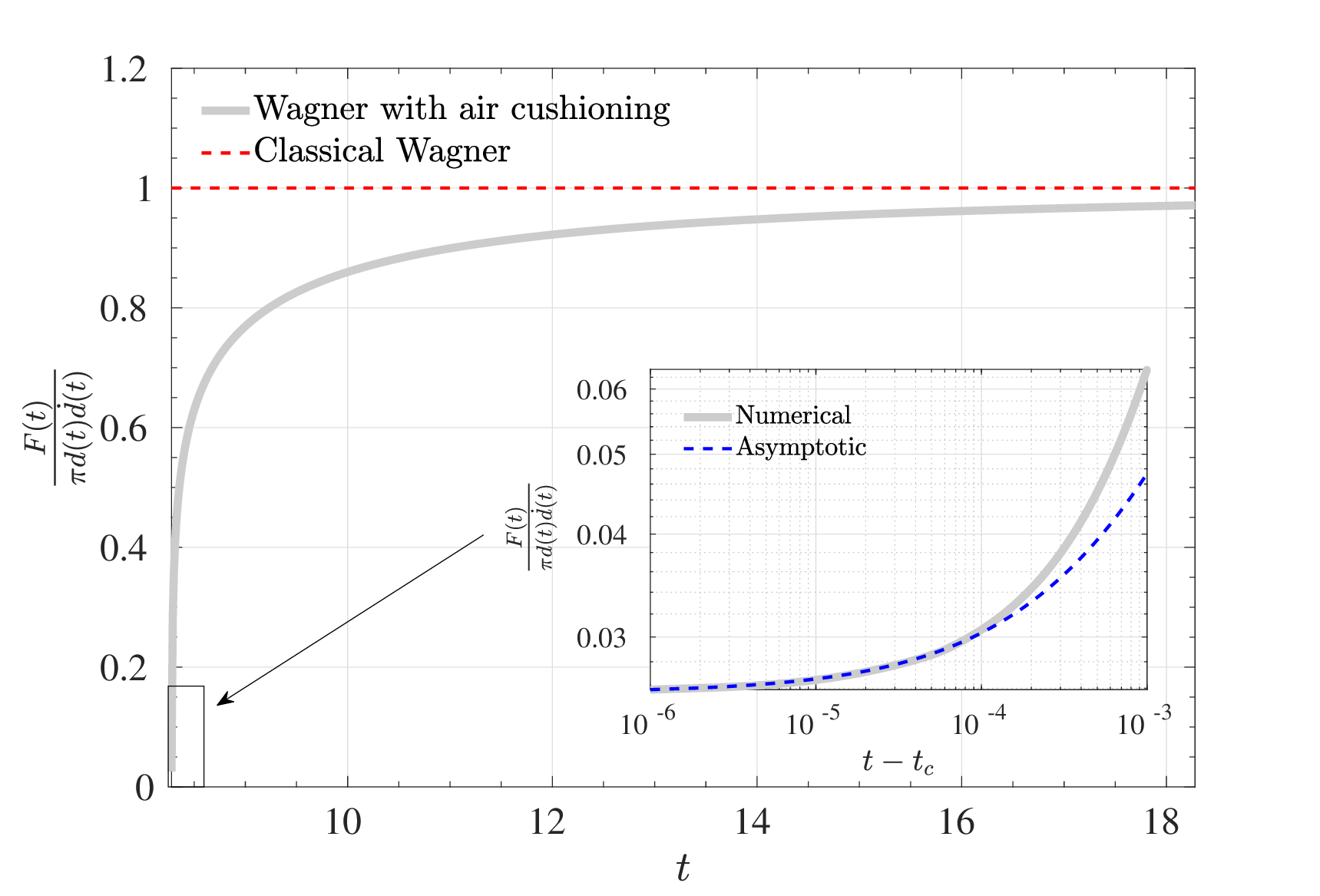}}
\caption{The leading-order force $F(t)$ on an impacting parabola scaled by $\pi d(t)\dot{d}(t)$. When there is no pre-impact air cushioning, the scaled force is simply $1$, as indicated by the dashed red line. The inset shows a log-log plot of the scaled force just after touchdown alongside the asymptotic solution (\ref{eqn:SmallTimeForce}) as a dashed blue line.}
\label{fig:Force}
\end{figure}

We can use this numerical solution for $d(t)$ to evaluate the scaled leading-order force on the impactor $F(t)/\pi d(t)\dot{d}(t)$ using (\ref{eqn:Force}), which we plot in figure \ref{fig:Force}. Note that, in the absence of air cushioning, this should simply be unity for all $t$, which is included as the red dashed line in the figure and, as expected from the far-field analysis of \textsection \ref{sec:LargeTimeSoln}, we approach this limit from below as $t\rightarrow\infty$. However, it is clear that the air has significantly reduced the slamming force when $t = O(1)$. In the inset, we compare the scaled leading-order force to the asymptotic expansion (\ref{eqn:SmallTimeForce}), showing excellent agreement. It is worth noting that, for this example, the numerical data for $v_{c}(x)$ for $-d_{c}<x<d_{c}$ is such that
\begin{linenomath}
\begin{equation}
 1 + \frac{1}{\pi}\int_{-1}^{1}\frac{v_{c}(d_{c}s)}{\sqrt{1-s^{2}}}\,\mbox{d}s \approx 0.026,
 \label{eqn:SmallTimeForceParabIntegral}
\end{equation}
\end{linenomath}
so that the force is singular at touchdown. Note that this singularity is not as apparent as it might be in figure \ref{fig:Force} since, for $t-t_{c} = O(10^{-6})$, $\tilde{F}_{0}$ given in (\ref{eqn:SmallTimeForce}a) is only $O(1)$ by virtue of (\ref{eqn:SmallTimeForceParabIntegral}). Naturally, improved fidelity of the pre-impact simulations would reduce this contribution even further, but comes with associated additional costs to computational time. Moreover, the points discussed previously in regards to the assumption (\ref{eqn:vcassumption}) still stand.

\begin{figure}[t!]
\centering \scalebox{0.4}{\epsfig{file=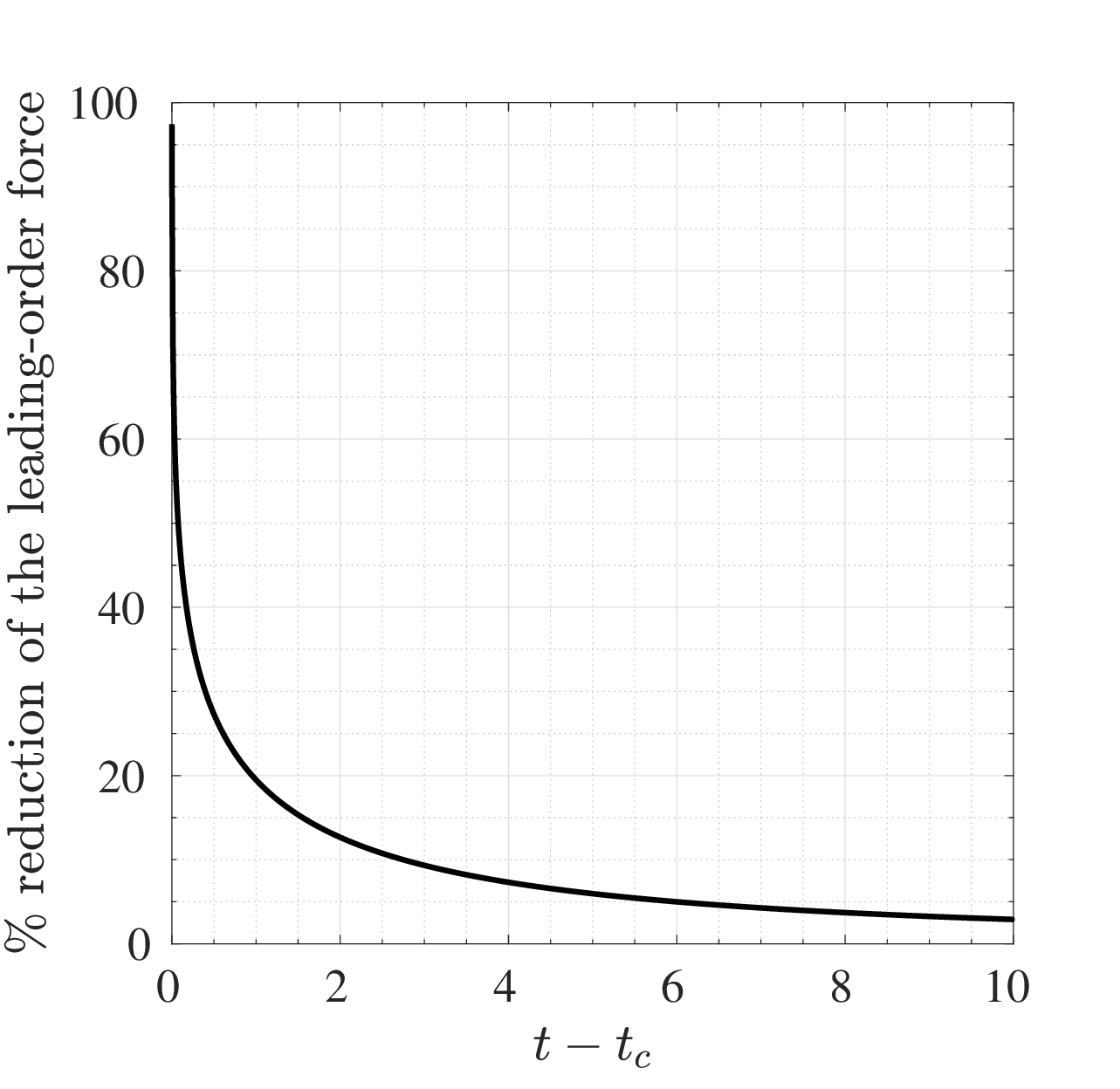}}
\caption{The percentage reduction of the scaled force on the body $F(t) / \pi d(t)\dot{d}(t)$ from the classical Wagner solution due to the pre-impact air-cushioning. }
\label{fig:ForceReduction}
\end{figure}

Throughout the impact, the effect of the air is to reduce the scaled force on the impactor compared to the classical Wagner prediction. We illustrate this in figure \ref{fig:ForceReduction}. The effect is most stark just after touchdown, where the reduction is $\approx 97.4\%$, but this has decreased to a $\approx 2.5\%$ reduction of the Wagner force at 10 dimensionless time units after touchdown. For reference, considering an example of a cylindrical body of radius $1$m impacting at $1$ m/s into a pool of water, the impact timescale $\ve^{2}R^{*}/V^{*} \approx 6.9\times10^{-6}$s. Touchdown  occurs at $t_{c}^{*} = \ve^{2}R^{*}t_{c}/V^{*} =  5.7 \times 10^{-5}$s, and we reach the final point in figure \ref{fig:ForceReduction} at $t^{*} = 1.3\times10^{-4}$s. Therefore the time period over which air cushioning has an appreciable effect on the (scaled) force is significant. However, we note that the ratio of the penetration depth to the radius of the cylinder at this time is thus $O(10^{-4})$, so this is still the early stage of entry.

\begin{figure}[t!]
\centering \scalebox{0.6}{\epsfig{file=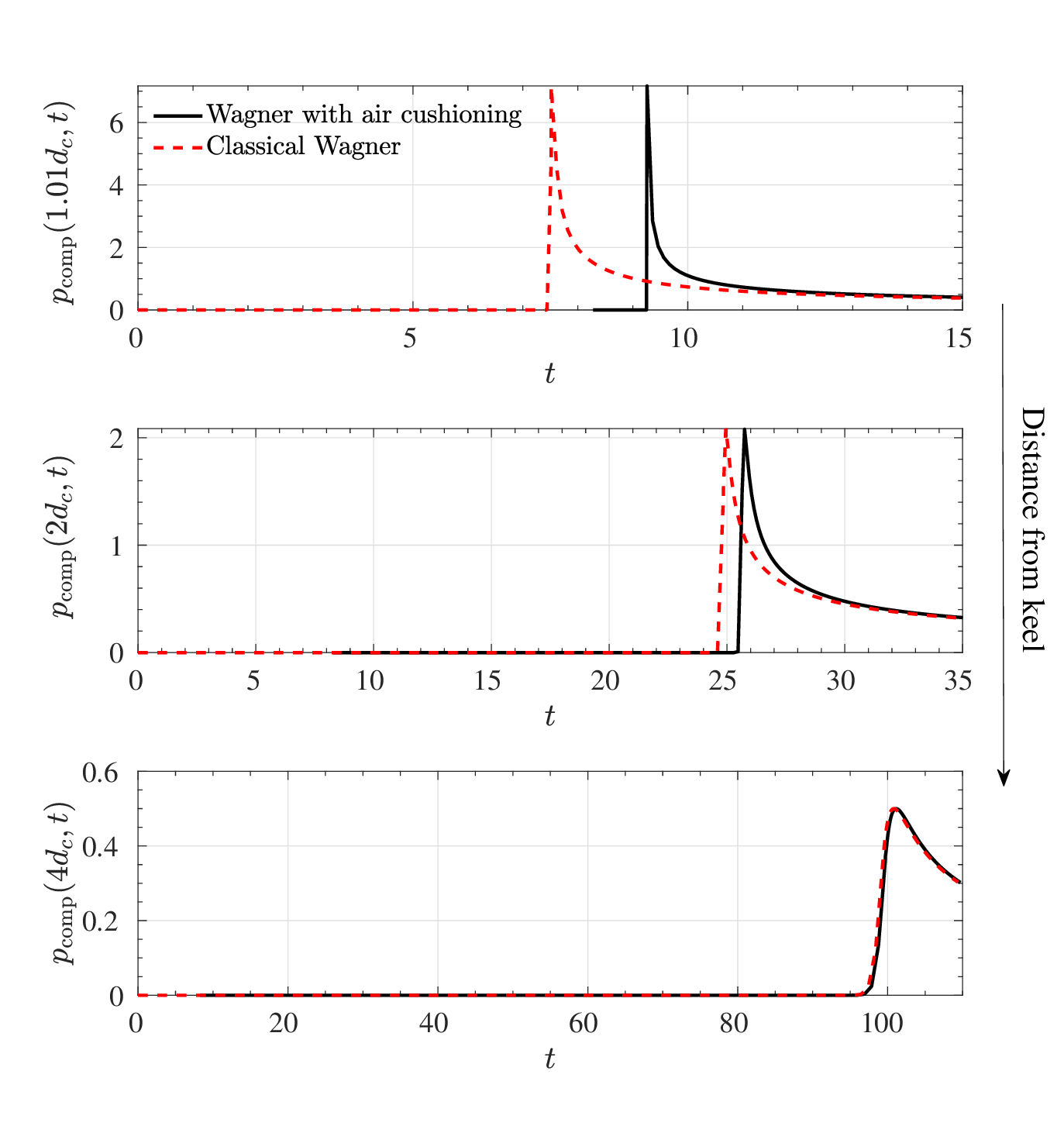}}
\caption{Composite pressure profiles on an impacting parabola at (from top to bottom) $x = 1.01d_{c}, 2d_{c}, 4d_{c}$. In each plot, the solid black line gives the composite pressure including the pre-impact air effects, while the red dashed line gives the equivalent composite pressure in the classical Wagner problem (when touchdown occurs at $t = 0$ at $x = 0$).}
\label{fig:PressureTimeSeries}
\end{figure}

In figure \ref{fig:PressureTimeSeries}, we plot time series of the composite pressure profile at different points on the body. Using van Dyke's matching rule, \cite{VanDyke1964}, the composite pressure is given by
\begin{linenomath}
 \begin{equation}
  p_{\mathrm{comp}} = p(x,t) + \frac{1}{\ve}\hat{p}((x-d(t))/\ve^{2},t) - p_{\mathrm{overlap}}(x,t),
 \end{equation}
\end{linenomath}
where $p(x,t)$ is given by (\ref{eqn:OuterPressure}), the jet-root pressure $\hat{p}(\hat{x},t)$ is given by (\ref{eqn:InnerPressure}) and $p_{\mathrm{overlap}}(x,t)$ is given by (\ref{eqn:OuterPressureExpansion}). In figure \ref{fig:PressureTimeSeries}, we have chosen three stations ahead of the initial touchdown point, $x = 1.01d_{c}, 2d_{c}, 4d_{c}$, and we show the evolution of the pressure on the impactor at each point. After impact, initially the pressure felt on that part of the body is very low, as its sole contribution would be from the splash jet, which is orders of magnitude smaller than the pressures in the outer and inner regions, \cite{Howison1991}. Once the jet-root region reaches the station, there is a rapid increase in pressure, the well-known Wagner pressure peak, \cite{Nethercote1986}, and then a slow fall as the jet-root moves further along the body. Clearly, at stations further from the keel of the impactor, this pressure peak is much lower, as the violence of the impact has diminished.  

To compare to the classical theory, we also plot the equivalent composite pressure profiles at the same time stations in the absence of any pre-impact air effects as the red dashed curves in figure \ref{fig:PressureTimeSeries}. The pressure profiles with and without air cushioning are remarkably similar at each station, although the air introduces a pronounced shift to later times. This suggests that the primary effect of air cushioning \emph{ahead of the touchdown point} is simply to delay the pressure response at a given location on the impactor, although as is clearly seen from the profile at $x = 4d_{c}$, this delay rapidly diminishes as we move further from the keel. 

We note that we have not, however, said anything about the pressure profiles closer to the keel than the initial point of touchdown. At these points, the largest pressures will be felt prior to touchdown, as the peaks in the free surface profile induced by the air-water interactions governed by (\ref{eqn:PreImpact1}), (\ref{eqn:PreImpact2}) lead to large air pressures, and hence a significant contribution to the force felt by the impacting body. Since it is the goal of the present analysis to concentrate on the role of the pre-impact air effects in the post-impact dynamics, we do not pursue this any further here, but this is likely to be the reason that classical Wagner theory significantly overestimates the pressure peaks very close to the keel of impacting bodies, but does extremely well at capturing the pressure further away, \cite{Howison1991}.

\begin{figure}[t!]
\centering \scalebox{0.5}{\epsfig{file=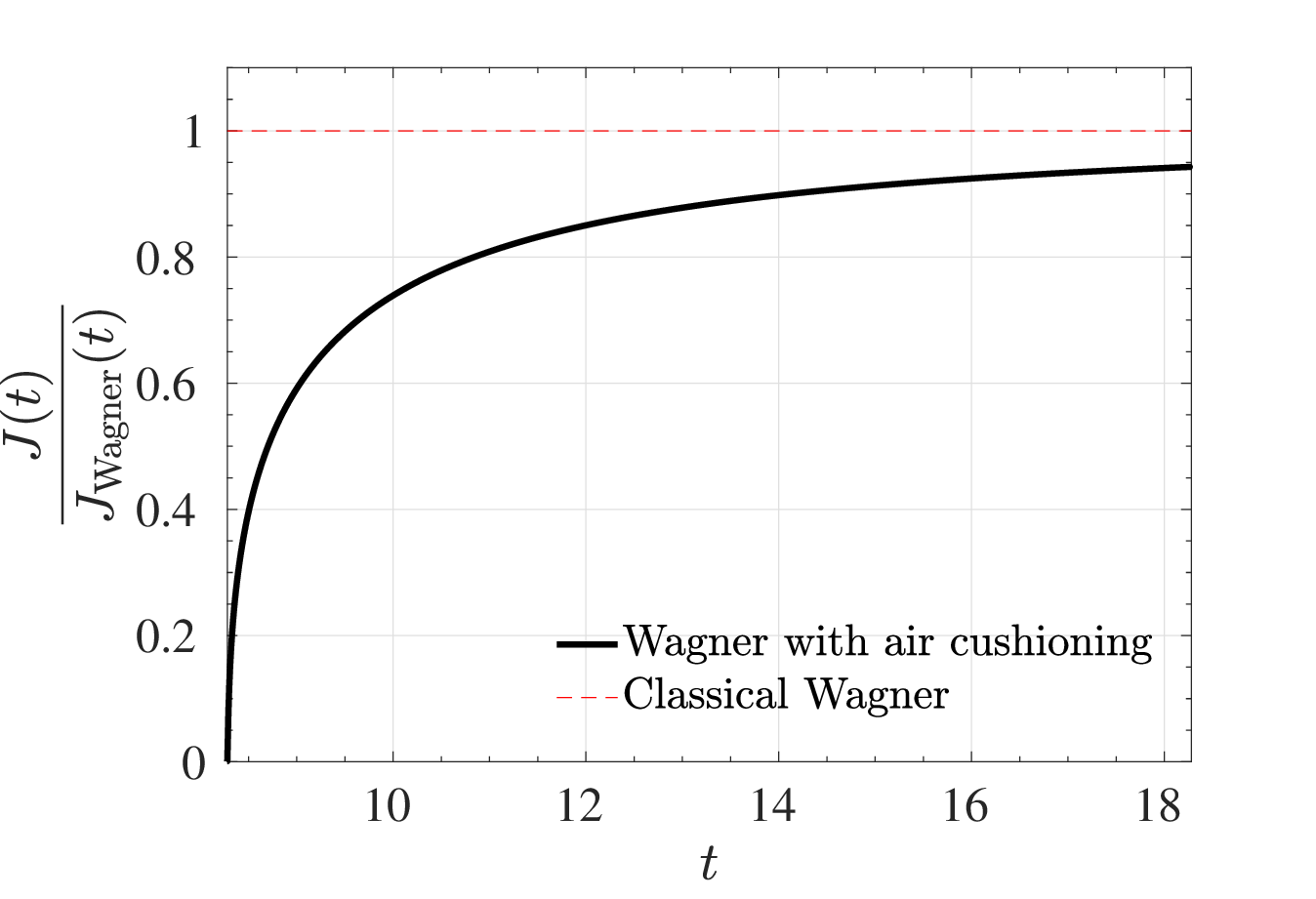}}
\caption{The reduction of the asymptotic jet thickness, $J(t)$, from the classical Wagner solution, $J_{\mathrm{Wagner}}(t) = \pi d(t)/8\dot{d}(t)^{2}$ as a function of time. Similar to the slamming force, at early times there is a significant reduction in the amount of fluid entering the jet, but this converges to the classical solution as the impact progresses.}
\label{fig:JetThickness}
\end{figure}

Finally, we consider the evolution of the jet thickness, $J(t)$, given by (\ref{eqn:JetThickness}). In figure \ref{fig:JetThickness}, we show $J(t)$ scaled by the Wagner jet thickness, $J_{\mathrm{Wagner}}(t) = \pi d(t)/8\dot{d}(t)^{2}$ to illustrate the role of the pre-impact air effects. On trend with the other impact properties, we see that at times just after touchdown, the jet thickness is substantially diminished from the Wagner solution, but as the impact progresses, we converge back to the classical solution. At early stages of the impact, this implies that the mass of fluid lost to the splash is thus smaller than the classical Wagner prediction, which may have direct relevance in applications where reducing fluid losses is important.

\section{Summary and discussion} \label{sec:Summary}

In this analysis, we have adapted Wagner theory for water-entry problems to account for the well-known effects of air cushioning before impact. Utilizing the analysis of \cite{Ross2019}, we showed that the air-water interactions that take place when the air gap between the solid impactor and the liquid pool is sufficiently thin can be incorporated into a Wagner-type theory post touchdown by neglecting the collapse and break-up of the entrapped air bubble. This assumption is justified in the sense that the air bubble is, to a reasonable approximation, moving down vertically at the same speed of the impactor close to touchdown and the timescale for bubble retraction and break up is much longer than the impact timescale.

Under this assumption, we formulated the post-impact problem as an adapted Wagner model in which the pre-impact air effects enter as initial conditions. We were able to derive expressions for the location of the jet roots, $x = \pm d(t)$, which give the size of the effective flat plate of the Wagner outer region, and the leading-order slamming force on the impactor. In each case, the effect of the air is to cushion the impact in comparison to the model in which the air is neglected, namely by slowing the growth of $d(t)$, reducing the force on the impactor and reducing the amount of fluid lost into the splash jets.

However, we were able to show asymptotically that the effects of the pre-impact free surface deformation and the resulting induced liquid flow become less important as the impact progresses, with the values of the turnover point location, the force on the impactor and the jet thickness at its base approaching the Wagner solution as $t\rightarrow\infty$. This partly explains the success of Wagner theory in predicting elements of droplet impact, as discussed by, for example, \cite{Philippi2016, Cimpeanu2018}, despite the absence of the air-cushioning behaviour in the classical model.

We also considered the role of the pre-impact dynamics on the pressure history at points on an impacting parabolic profile ahead of the initial point of touchdown. While the air-cushioning delays the pressure response felt at a particular location, the actual pressure profiles are remarkably similar to the classical Wagner model. Moreover, the delay decreases rapidly as we move further from the keel of the impactor, indicating, again, that as the impact progresses the effect of the air cushioning dies away, and the Wagner solution does admirably in capturing the later-stage impact dynamics.

Hence, in all of our analysis, we have shown that pre-impact air-cushioning has an important influence just after touchdown, but that this influence diminishes as time increases. These findings may be of significance in industries where the primary interest is in effects very close to touchdown/entry, for example in ship-slamming or wave impacts on coastal structures, where the peak stresses are of interest, \cite{Wang2017, Ghadirian2019}. Our analysis has shown that classical Wagner theory significantly over-estimates, for example, the extent of the effective contact set and the slamming force at these times. However, if the important effects are after the initial stage of slamming has died away, our results show that neglecting the effect of the pre-impact air-cushioning is a reasonable assumption.

It would be an interesting extension of the current analysis to include a model for the entrapped air bubble after touchdown in the post-impact dynamics in order to rigorously scrutinize the assumption made in the present analysis that its effect is small. One possible avenue for this may be to model the bubble as a patch cavity, as in for example, \cite{Howison1994, Korobkin2003, Moore2012}. The slow timescale of bubble retraction may aid such an analysis, although care must be taken in specifying suitable conditions at the trailing edge of the bubble.

Finally, we note that the post-impact methodology does not require the use of the pre-impact results of the reduced coupled air-water model as described in \textsection \ref{sec:PreImpact} as the initial conditions for the post-impact model. Indeed, given the power of direct numerical simulations (DNS) of the full Navier-Stokes equations in studying impact dynamics, \cite{Thoraval2012, Cimpeanu2018}, one could readily take data from DNS of a given water-entry problem to find the time and location of touchdown and the appropriate initial profiles for the free surface, free surface velocity and velocity potential. This would also offer the exciting possibility of extensively comparing DNS to the predictions of the adapted Wagner model to study the importance of pre-impact air effects in predicting properties of the post-impact dynamics, such as the location of the jet root and the jet speed and thickness along its entire length.

\section*{Acknowledgments}

I would like to thank the comments of the anonymous referees, which helped improve upon a previous version of this paper.

%%%%%%%%%%%%%%%%%%%%%%%%%%%%%%%%%%%%%%%%%%%%%%%%%%%%%%%%%%%%%%%%%%%%%%%%%%%%
%%%%%%%%%%%%%%%%%%%%%%%%%%%%%%%%%%%%%%%%%%%%%%%%%%%%%%%%%%%%%%%%%%%%%%%%%%%%
%%%%%%%%%%%%%%%%%%%%%%%%%%%%%%%% APPENDICES %%%%%%%%%%%%%%%%%%%%%%%%%%%%%%%%
%%%%%%%%%%%%%%%%%%%%%%%%%%%%%%%%%%%%%%%%%%%%%%%%%%%%%%%%%%%%%%%%%%%%%%%%%%%%
%%%%%%%%%%%%%%%%%%%%%%%%%%%%%%%%%%%%%%%%%%%%%%%%%%%%%%%%%%%%%%%%%%%%%%%%%%%%

\appendix

\section{Jet-root region}
\label{sec:AppendixWagnerInner}

The leading-order linearized post-impact problem in figure \ref{fig:PostImpactOuterMBVP} predicts velocities and pressures that are square-root singular as $x\rightarrow\pm d(t)$. Naturally, the asymptotic model breaks down close to these points, and there is an inner region in which nonlinear terms in the free surface boundary conditions become important. Returning to $\phi$ rather than $\Phi$, the full outer region problem is given by
\refstepcounter{equation}
\begin{linenomath}
 $$
 \nabla^{2}\phi = 0, \quad \frac{\partial\phi}{\partial t} + p + \frac{\ve}{2}|\nabla\phi|^{2} = 0 \quad \mbox{in the fluid},
 \label{eqn:FullOuter1}
 \eqno{(\theequation{\mathit{a},\mathit{b}})}
 $$
\end{linenomath}
subject to
\begin{linenomath}
 \begin{equation}
  \frac{\partial\phi}{\partial y} = -1 - \ve f'(x)\frac{\partial\phi}{\partial x}
  \label{eqn:FullOuter2}
 \end{equation}
\end{linenomath}
on $y = \ve(f(x)-t)$, $|x|<d(t)$, and
\refstepcounter{equation}
\begin{linenomath}
$$
p = 0, \quad \frac{\partial\phi}{\partial y} = \frac{\partial h}{\partial t} + \ve\frac{\partial \phi}{\partial x}\frac{\partial h}{\partial x}
\label{eqn:FullOuter3}
\eqno{(\theequation{\mathit{a},\mathit{b}})}
$$
\end{linenomath}
on $y = \ve h(x,t)$, $|x|>d(t)$. Thus in order to bring in the nonlinearities local to the turnover points, the inner or \emph{jet-root} region is two orders of magnitude smaller than the linearized region, \cite{Howison1991}, so that the pertinent scalings are
\begin{linenomath}
\begin{equation}
 x = d(t) + \ve^{2}\hat{x}, \quad y = \ve\left(f(d(t))-t\right) + \ve^{2}\hat{y}, \quad h = f(d(t))-t + \ve\hat{h}, \quad \phi = \ve\left(\dot{d}(t)\hat{x} + \hat{\phi}\right), \quad p = \frac{\hat{p}}{\ve}.
\end{equation}
\end{linenomath}

Under these scalings, the jet-root problem is thus given by
\refstepcounter{equation}
\begin{linenomath}
 $$
 \nabla^{2}\hat{\phi} = 0, \quad \hat{p} - \frac{\dot{d}(t)^{2}}{2} + \frac{1}{2}|\nabla\hat{\phi}|^{2} = \ve \frac{\mbox{d}}{\mbox{d}t}\left(f(d(t))-t\right)\frac{\partial\hat{\phi}}{\partial\hat{y}}-\ve^{2}\left(\ddot{d}(t)+\frac{\partial\hat{\phi}}{\partial t}\right) \quad \mbox{in the fluid},
 \label{eqn:FullInner1}
 \eqno{(\theequation{\mathit{a},\mathit{b}})}
 $$
\end{linenomath}
subject to
\begin{linenomath}
 \begin{equation}
  \frac{\partial\hat{\phi}}{\partial\hat{y}} = \ve\left(-1-f'(d(t) + \ve^{2}\hat{x})\right)\left(\dot{d}(t)+\frac{\partial\hat{\phi}}{\partial\hat{x}}\right)
   \label{eqn:FullInner2}
 \end{equation}
\end{linenomath}
on the body, which is locally given by $\hat{y} = \ve f'(d(t))\hat{x} + O(\ve^{2})$, and
\refstepcounter{equation}
\begin{linenomath}
 $$
 \hat{p} = 0, \quad \frac{\partial\hat{\phi}}{\partial\hat{y}} = \frac{\partial\hat{h}}{\partial\hat{x}}\frac{\partial\hat{\phi}}{\partial\hat{x}} + \ve\frac{\mbox{d}}{\mbox{d}t}\left(f(d(t))-t\right) + \ve^{2}\frac{\partial\hat{h}}{\partial t} \label{eqn:FullInner3}
  \eqno{(\theequation{\mathit{a},\mathit{b}})}
 $$
\end{linenomath}
on $\hat{y} = \hat{h}$. This is supplemented by the matching condition in the far-field,
\begin{linenomath}
 \begin{equation}
  \frac{\partial\hat{\phi}}{\partial\hat{x}} - \mbox{i}\frac{\partial\hat{\phi}}{\partial\hat{y}} \sim -\dot{d}(t) + \frac{S(t)}{\sqrt{\hat{x}^{2}+\hat{y}^{2}}} \quad \mbox{as} \quad \hat{x}^{2}+\hat{y}^{2}\rightarrow\infty,
  \label{eqn:FullInner4}
 \end{equation}
\end{linenomath}
where $S(t)$ is the coefficient of the inverse square-root singularity in the leading-order-outer velocity.

It is clear that, to leading order in $\ve$, the inner problem (\ref{eqn:FullInner1})--(\ref{eqn:FullInner3}) is unchanged from the standard Wagner inner region, \cite{Howison1991}. Thus, the leading-order-inner solution is given parametrically by
\begin{linenomath}
 \begin{equation}
  \hat{\phi} + \mbox{i}\hat{\psi} = C(t) + \frac{\dot{d(t)}J(t)}{\pi}\left(\zeta-\log\zeta\right), \quad \hat{x} + \mbox{i}\hat{y} = -\frac{J(t)}{\pi}\left(1+\zeta +4\sqrt{\zeta} + \log\zeta\right),
 \end{equation}
\end{linenomath}
where the fluid domain is in $\mbox{Im}(\zeta)>0$, the branch cuts are taken along the negative real-$\zeta$ axis and the free surface lies on $\zeta = \xi + \mbox{i}0$, $\xi<0$, \cite{Wagner1932}. The unknown function $C(t)$ may only be determined by proceeding to higher order in the asymptotic analysis, but it is not required for any of the predictions we make in the leading-order model. The function $J(t)$ gives the leading-order jet thickness as we move out of the inner region, and is found by matching the inner and outer fluid velocities to be
\begin{linenomath}
 \begin{equation}
  J(t) = \frac{\pi S(t)^{2}}{16\dot{d}(t)^{2}}.
 \end{equation}
\end{linenomath}
Note that the jet thickness at the jet-root depends on the pre-impact air-cushioning
model through both $d(t)$, which satisfies the Wagner condition (\ref{eqn:d}) and through $S(t)$. By expanding (\ref{eqn:LOOComplexVelocity}) close to the contact point, we deduce that
\begin{linenomath}
 \begin{equation}
  J(t) = \frac{\pi d(t)}{8\dot{d}(t)^{2}}\left(1+\frac{1}{\pi}\int_{-1}^{1}\sqrt{\frac{1+S}{1-S}}v_{c}(d(t)S)\,\mbox{d}S\right)^{2}.
 \end{equation}
\end{linenomath}
 
Finally, the leading-order-inner pressure on the impactor is given parametrically by
\begin{linenomath}
 \begin{equation}
  \hat{x} = \frac{-J(t)}{\pi}\left(1+\xi+4\sqrt{\xi}+\log\xi\right), \quad \hat{p}(\xi) = \frac{\dot{d}(t)^{2}}{2}\left(1-\left(\frac{1-\sqrt{\xi}}{1+\sqrt{\xi}}\right)^{2}\right), 
  \label{eqn:InnerPressure}
 \end{equation}
\end{linenomath}
where $\xi >0$.

\section{Numerical methodology for inverting the Wagner condition} \label{sec:AppendixNumericalTurnoverInversion}

It is most straightforward to invert the Wagner condition (\ref{eqn:d}) by making the assumption that $x = d(t)$ is invertible, and we define
\begin{linenomath}
 \begin{equation}
  x = d(t) \Longleftrightarrow t = \omega(x).
 \end{equation}
\end{linenomath}
Note that this assumption is valid provided $\dot{d}(t)>0$, which is also a necessary condition for stability of the classical Wagner problem, \cite{Howison1991}, and is typically reasonable for problems in which the impact velocity is vertical and uniform, but may need closer investigation if the impact velocity is oblique, \cite{Moore2012}, or if the solid body is taken as deformable, \cite{Pegg2018}. While the methodology is readily generalized, we shall present it for the case of an impacting parabola, as this is most relevant to the current study.

If we substitute $t = \omega(x)$ into (\ref{eqn:d}), we find
\begin{linenomath}
 \begin{alignat}{2}
  \frac{x^{2}}{2} - t_{c} - h_{c}(x) - v_{c}(x)(\omega-t_{c}) & = && \, \int_{d_{c}}^{x}\frac{\omega'(\sigma)x}{\sqrt{x^{2}-\sigma^{2}}}\,\mbox{d}\sigma - \frac{1}{\pi}
  \int_{d_{c}}^{x}\frac{\omega'(\sigma)\sigma^{2}}{\sqrt{x^{2}-\sigma^{2}}}\int_{-1}^{1}\frac{\sqrt{1-\xi^{2}}}{\sigma\xi-x}v_{c}(\sigma\xi)\,\mbox{d}\xi\mbox{d}\sigma. 
  \label{eqn:AppwWagner}
 \end{alignat}
\end{linenomath}
Then, starting with the small-time asymptotic solution (\ref{eqn:SmallTimed}), we can solve for $\omega(x)$ by marching forward in $x$ as follows. Suppose that we know the solution up to a station $x = x^{\dag}-\Delta x$. Then, at $x^{\dag}$, we have
\begin{linenomath}
 \begin{alignat}{2}
  & && \,\int_{x^{\dag}-\Delta x}^{x^{\dag}} \omega'(\sigma)\left(\frac{x}{\sqrt{x^{2}-\sigma^{2}}}-\frac{1}{\pi}\frac{\sigma^{2}}{\sqrt{x^{2}-\sigma^{2}}}\int_{-1}^{1}\frac{\sqrt{1-\xi^{2}}}{\sigma\xi-x}v_{c}(\sigma\xi)\,\mbox{d}\xi\right)\,\mbox{d}\sigma + v_{c}(x)w(x^{\dag}) = \frac{x^{2}}{2} - t_{c} - h_{c}(x) + \nonumber \\
  & && \qquad \qquad v_{c}(x)t_{c}-\int_{d_{c}}^{x^{\dag}-\Delta x} \omega'(\sigma)\left(\frac{x}{\sqrt{x^{2}-\sigma^{2}}}-\frac{1}{\pi}\frac{\sigma^{2}}{\sqrt{x^{2}-\sigma^{2}}}\int_{-1}^{1}\frac{\sqrt{1-\xi^{2}}}{\sigma\xi-x}v_{c}(\sigma\xi)\,\mbox{d}\xi\right)\,\mbox{d}\sigma + v_{c}(x)w(x^{\dag}).
 \end{alignat}
\end{linenomath}
Everything on the right-hand side is known, while the first integral on the left-hand side can be approximated asymptotically as
\begin{linenomath}
 \begin{alignat}{2}
  & && \, \int_{x^{\dag}-\Delta x}^{x^{\dag}} \omega'(\sigma)\left(\frac{x}{\sqrt{x^{2}-\sigma^{2}}}-\frac{1}{\pi}\frac{\sigma^{2}}{\sqrt{x^{2}-\sigma^{2}}}\int_{-1}^{1}\frac{\sqrt{1-\xi^{2}}}{\sigma\xi-x}v_{c}(\sigma\xi)\,\mbox{d}\xi\right)\,\mbox{d}\sigma  \nonumber \\
  & && \quad = \omega'(x^{\dag})\sqrt{2x^{\dag}}\left(1 + \frac{1}{\pi}\int_{-1}^{1}\sqrt{\frac{1+\xi}{1-\xi}}v_{c}(x^{\dag}\xi)\,\mbox{d}\xi\right)\sqrt{\Delta x} - \omega'(x^{\dag})v_{c}(x^{\dag})\Delta x + o(\Delta x).
 \end{alignat}
\end{linenomath}
It is then straightforward to solve for $\omega(x^{\dag})$.

%%%%%%%%%%%%%%%%%%%%%%%%%%%%%%%%%%%%%%%%%%%%%%%%%%%%%%%%%%%%%%%%%%%%%%%%%%%%
%%%%%%%%%%%%%%%%%%%%%%%%%%%%%%%%%%%%%%%%%%%%%%%%%%%%%%%%%%%%%%%%%%%%%%%%%%%%
%%%%%%%%%%%%%%%%%%%%%%%%%%%%%%%% REFERENCES %%%%%%%%%%%%%%%%%%%%%%%%%%%%%%%%
%%%%%%%%%%%%%%%%%%%%%%%%%%%%%%%%%%%%%%%%%%%%%%%%%%%%%%%%%%%%%%%%%%%%%%%%%%%%
%%%%%%%%%%%%%%%%%%%%%%%%%%%%%%%%%%%%%%%%%%%%%%%%%%%%%%%%%%%%%%%%%%%%%%%%%%%%

\nocite{*}

% BibTeX users please use one of
% \bibliographystyle{spbasic}      % basic style, author-year citations
\bibliographystyle{spmpsci}      % mathematics and physical sciences
\bibliography{NonQuiescentWagner.bib}  % name your BibTeX data base

\end{document}